\documentstyle[prl,aps,amssymb,graphicx]{revtex}
\newcommand{\beq}{\begin{equation}}
\newcommand{\eneq}{\end{equation}}
\begin{document}

\tolerance 10000

\twocolumn[\hsize\textwidth\columnwidth\hsize\csname %
@twocolumnfalse\endcsname

\draft

\title{Coordinate Representation of the One-Spinon One-Holon
Wavefunction and Spinon-Holon Interaction}

\author {B. A. Bernevig$~^{*,+}$, D. Giuliano$~^{\dagger,\%}$ and R. B.
         Laughlin $~^*$}

\address{$~^*$Department of Physics, Stanford University,
        Stanford, California 94305\\
        $~^\dagger$ Istituto Nazionale di Fisica della Materia (INFM),
         Unit\`a di Napoli, Napoli, Italy\\
         $^\%$ Dipartimento di Scienze Fisiche Universit\`a di Napoli
         "Federico II ",\\
         Monte S.Angelo - via Cintia, I-80126 Napoli, Italy\\
        $^{+}$ Department of Physics, Massachusetts Institute of Technology,
        Cambridge, MA 02139 }

\date{\today}
\maketitle
\widetext

\begin{abstract}
\begin{center}
\parbox{14cm}{By deriving and studying the coordinate representation for the
one-spinon one-holon wavefunction we show that spinons and holons in the
supersymmetric $t - J$ model with $1/r^2$ interaction attract each other. The
interaction causes a probability enhancement in the one-spinon one-holon
wavefunction at short separation between the particles. We express the hole
spectral function for a finite lattice in terms of the probability enhancement,
given by the one-spinon one-holon wavefunction at zero separation. In the
thermodynamic limit, the spinon-holon attraction turns into the square-root
divergence in the hole spectral function.}

\end{center}
\end{abstract}

\pacs{
\hspace{1.9cm}
PACS numbers: 75.10.Jm, 71.27.+a, 05.30.Pr
}
]

\narrowtext

\section{Introduction}

Landau's Fermi liquid theory applies to interacting electron systems that can
be adiabatically deformed to a Fermi gas. If the interaction is
smoothly switched off, the spectrum of a Fermi liquid reduces to the
spectrum of a noninteracting fermionic system. The excitations
of a Fermi liquid are given by quasiparticles and quasiholes. Although
their lifetime may be short,
it always becomes infinite at the Fermi surface \cite{landau}. 
As  one is concerned only with energies close to the Fermi surface, Fermi
liquid's picture applies to a wide class of correlated systems. Experimentally,
Landau quasiparticles are observed as a resonant peak at the Fermi surface 
in the spectral density of states measured at fixed momentum.

Nevertheless, there are several low-dimensional strongly correlated
systems where Landau's picture breaks down. In 
Luttinger liquids the spin and
charge degrees of freedom ``separate'' and the quasiparticles and quasiholes
are no longer elementary excitations \cite{luttinger,several}. The same 
phenomenon was discovered by means of Bethe-ansatz-like techniques in 
exactly solvable models, like the supersymmetric $t-J$-model with 
$1/r^2$-interaction, which we study in this paper \cite{kurakato}.

Physically a particle or a hole injected in a strongly correlated chain breaks
up into particles carrying spin, but no charge -spinons- and charge, but 
no spin - holons. The quantum numbers of particles and
holes ``fractionalize''. Spinons and holons are the true 
elementary excitations of the strongly correlated chains \cite{spho,haha}.
When widely separated one from each other, spinons
and holons propagate independently, in general with different velocities.
Such a phenomenon is usually referred to as ``spin-charge separation''.
As a consequence of spin-charge separation, the quasiparticle peak disappears
and is substituted by a broad spectrum. This was experimentally
detected in ARPES experiments on quasi-1D samples \cite{zxshen}.

Using Bethe-Ansatz solutions of 1D models, one can work out the energy of
a many-spinon many-holon state. In the thermodynamic limit the total energy
is the sum of the energies of each isolated particle. However, this does not
imply that spinons and holons do not interact. In recent papers
\cite{us1,us2} we carefully studied spinon interaction in an exact solution of
the Haldane-Shastry model (HSM) \cite{haldane,shastry}, a model for a
strongly correlated 1D system with no charge
degrees of freedom. We showed that spinons
interact, although in the thermodynamic limit the energy of a many-spinon
solution is additive. We also showed that their interaction is responsible for
the lack of integrity of spin waves against decay into spinons, the true
low-energy excitations of the system.

In this paper we generalize the formalism introduced in
Ref.\cite{us2} to study the
interaction between spinons and holons in an exact closed-form solution of
the supersymmetric extension of the HSM: the supersymmetric $t-J$-model
with $1/r^2$-interaction (Kuramoto-Yokoyama (KY) model
\cite{kuramoto}.)  The KY-model is a system of electrons
located at the sites of a circular lattice, where double occupancy of
a site is forbidden by strong Coulomb repulsion. Charge hopping,
Coulomb interaction and spin-spin antiferromagnetic interaction are all
inversely proportional to the square of the chord between the corresponding
sites. Charge vacancies at some sites (holes) may be created by filling the
system with fewer electrons than the number of sites. In the KYM one is
concerned with both spin and charge degrees of freedom.

We investigate the basic features of the KY-model by employing a formalism
based on analytic variables on the unit radius circle. As in the case
of the HSM, it easier to construct
and visualize the spinon and holon excitation by using a real-space
formalism than by using the Bethe-Ansatz
formalism. Within our formalism we derive a
``real space'' representation of the one-spinon one-holon wavefunction as
a solution of an appropriate equation of motion.

By taking the thermodynamic limit of
the exact solution of the equation of motion, we find that the
probability of finding a spinon and a holon at large separations from each 
other is independent of the distance between the particles, while it 
is greatly enhanced when the two particles are on top of each other, 
phenomenon we refer to as ``short-distance spinon-holon probability 
enhancement''. In the thermodynamic
limit, the spinon-holon interaction assumes the same form as the spinon-spinon
interaction derived in \cite{us1}, although the corresponding equations of
motion are completely different. The physical interpretation is the same
as in the case of the spinon-spinon interaction: a spinon and a holon do not
interact when they are widely separated from each other, while they
exhibit a short-range attraction at short separations \cite{us3}.

To show the effects of spinon-holon interaction on
the hole spectral density, $A_{\rm h} (\omega, q)$, we exactly calculate 
the contribution to $A_{\rm h} (\omega, q)$ from one-spinon one-holon states,
$A_{\rm h}^{\rm sp,ho} (\omega , q)$. The spinon-holon interaction has
important consequences on the functional form of $A_{\rm h}^{\rm sp,ho}
(\omega, q)$. The probability enhancement causes the overlap between the
wavefunction for the localized
hole and that for a spinon-holon pair to be significant, although not enough
to form a spinon-holon bound state. The corresponding matrix element is
enhanced - despite the fact that the density of states is uniform at low
energy - so as to make the hole excitation fully unstable to decay into a
spinon-holon pair.  Taking the thermodynamic limit of our result, we show
that, as the size of the system increases, spinon-holon interaction turns
into a square root singularity at the one-spinon one-holon creation threshold.
Correspondingly,  $A_{\rm h}^{\rm sp,ho} (\omega, q)$ shows no Landau
quasiparticle's peak, but it rather exhibits a sharp singular threshold,
followed by a broad branch cut \cite{kato}.

The paper is organized as follows: In section II we shortly review the
KY Hamiltonian and its supersymmetry; In Section III we introduce the
ground state of the KY model at half filling and its representation as a
function of analytic variables on the unit circle. At half-filling, the
ground state is the same as the ground state of the HS-model -
a disordered spin singlet. We will briefly review some properties of the
ground state, already discussed at length in \cite{us2}; In section IV we
analyze the one-spinon solution and review its relevant properties; In
section V we focus on the one-holon solution and derive its relevant
properties. In section VI we derive the action of ${\cal H}_{KY}$
on the one-spinon one-holon states, the energy eigenvalues, the corresponding
eigenvectors, and their norm; in Section VII we write the
Schr\"odinger equation for the one-spinon one-holon wavefunction,
whose solutions are simple polynomials. From the behavior of the
one-spinon one-holon wavefunction, we infer the nature of the
interaction between spinons and holons: a short-range attraction.  The
physical consequences of such an interaction are discussed at length
in Section VIII, where we rederive an exact expression for the
contribution of one-spinon one-holon states to the hole spectral
function in terms of the spinon-holon wavefunctions and rigorously
prove that this contribution is completely determined by the spinon-holon
interaction. In the thermodynamic limit spinon-holon interaction turns
into the square root divergence in the hole spectral function,
obtained by \cite{kato}. In Section IX we provide our main conclusions.

\section{Kuramoto-Yokoyama Hamiltonian}

The Kuramoto-Yokoyama model is defined on a lattice with periodic
boundary conditions. Sites are parametrized by the $N$-th roots of unity,
 $z_\alpha$ ($\alpha = 1 , \ldots , N$.) Strong electron repulsion
forbids double occupancy at each site. Therefore, sites can be occupied
by an $\uparrow$ or a $\downarrow$-electron, or they can be empty. The number of
empty sites can be tuned by means of an external chemical potential which
fixes the total charge of the system. The Kuramoto-Yokoyama Hamiltonian
\cite{kuramoto} is a generalization of the Haldane-Shastry Hamiltonian
\cite{haldane,shastry}, where also charge dynamics is taken into account.
It takes the form:

\[
{\cal H}_{KY} = J \left( \frac{2\pi}{N} \right)^2
\sum_{\alpha < \beta}^N \frac{1}{ | z_\alpha - z_\beta |^2 } \;
P \biggl\{ \vec{S}_\alpha \cdot \vec{S}_\beta
\]

\begin{equation}
- \frac{1}{2} \sum_\sigma
( c_{\alpha \sigma}^\dagger c_{\beta \sigma}  )
+ \frac{1}{2} (n_\alpha + n_\beta) - \frac{1}{4} n_\alpha n_\beta
- \frac{3}{4} \biggr\} P \; ,
\label{KYHamiltonian}
\end{equation}
\noindent
where the Gutzwiller's projector

\begin{equation}
P = \prod_\alpha \; ( 1 -
c_{\alpha \uparrow}^\dagger c_{\alpha \downarrow}^\dagger
c_{\alpha \downarrow} c_{\alpha \uparrow} )
\; \; \; ,
\label{gutz}
\end{equation}
\noindent
accounts for the no-double occupancy constraint.

Site occupation and spin operators are given by

\begin{equation}
n_\alpha =  c_{\alpha \uparrow}^\dagger c_{\alpha \uparrow}
+ c_{\alpha \downarrow}^\dagger c_{\alpha \downarrow}
\label{nopeartor}
\end{equation}

\begin{equation}
S_\alpha^a = \frac{1}{2} \sum_{\sigma , \sigma^{'}}
c_{\alpha \sigma}^\dagger \tau^a_{\sigma \sigma^{'}} c_{\alpha \sigma^{'}}
\; \; \; .
\label{szoperator}
\end{equation}
\noindent
where $\tau^a$, $a=x,y,z$, are Pauli matrices.
An empty site corresponds to a charge-1 hole that can tunnel to nearby sites
by means of the same inverse-square matrix element characterizing the spin
exchange and the charge-repulsion term in ${\cal H}_{KY}$.

Usual bosonic symmetries of a $t-J$-like model are total spin, corresponding
to the operator $\vec{S} = \sum_\alpha \vec{S}_\alpha$, and total
charge, corresponding to the operator $N= \sum_\alpha n_\alpha$.
The equivalence of energy scales for magnetism, charge transport and
charge interaction, causes the KY Hamiltonian to be supersymmetric, in the
sense that it commutes with the electron or hole injection operators
${\cal Q}_\sigma  = \sum_\alpha P c_{\alpha \sigma} P$,
${\cal Q}_\sigma^\dagger  = \sum_\alpha P c_{\alpha \sigma}^\dagger P$.

As in the Haldane-Shastry hamiltonian,  since the complex variable $z$ 
lays on the unit circle ($z^* = z^{-1}$), the interaction is an
analytic function of the coordinates, that is:

\begin{displaymath}
\frac{1}{ | z_\alpha - z_\beta |^2} = - \frac{ z_\alpha z_\beta}{ ( z_\alpha -
z_\beta )^2} \;\;\; .
\end{displaymath}
\noindent
Throughout the paper we use the representation in terms of the analytic
variables $z_\alpha$ \cite{chia,us2}. This turns out to be very useful for 
describing the properties of spinons and holons in real space.

\section{Ground State}
In this section and in the following one we discuss the ground-state
and the one-spinon eigenstates of the KY-model at half filling.
At half-filling the KY-model reduces to the Haldane Shastry (HS)
Hamiltonian \cite{haldane,shastry}

\beq
{\cal H}_{HS} = J \left( \frac{2 \pi}{ N} \right)^2 \sum_{\alpha < \beta}^N
\frac{ \vec{S}_\alpha \cdot \vec{S}_\beta}{ | z_\alpha - z_\beta |^2} \;\;\; .
\eneq
\noindent
Both the ground state and the one-spinon
eigenstates are the same as for the HS-model.
Since in \cite{us2} we have already applied our formalism to study basic
properties of the HS-model, here we will only briefly review the main
results in view of their extension to states where holon excitations
are present.

\subsection{Ground State wavefunction}

Let $N$ be even. We first give the representation of the
ground state $ | \Psi_{GS} \rangle$ in terms of the $z$-coordinates and
then derive its energy.  $| \Psi_{GS} \rangle $ is defined in terms
of its projection onto the set of states with $M=N/2$ spins up and the
remaining spins down. If $z_1 , \ldots , z_{M}$ are the coordinates of
the up spins, one defines the state $ | z_1 , \ldots , z_{M} \rangle$
as: $ | z_1 , \ldots , z_{M} \rangle = \prod_{j=1}^M S_j^+
\prod_{\alpha = 1}^N c_{\alpha \downarrow}^\dagger | 0 \rangle$ where
$ | 0 \rangle$ is the empty state. The projections are given by:

\begin{equation}
\Psi_{GS} (z_1, \ldots , z_{M}) =
\prod_{j<k}^{M} (z_j - z_k)^2
\prod_{j=1}^{M} z_j  \;\;\; .
\label{gstate}
\end{equation}
\noindent
$\Psi_{GS}$ is a polynomial in the analytic variables $z_1 , \ldots , z_M$. Its
norm was first computed by Wilson  \cite{kwil}
by using the following identity:

\[
C_M = \sum_{ z_1 , \ldots , z_{M}} \prod_{i<j}^M
| z_i - z_j |^4
\]

\beq
= ( \frac{N}{2 \pi i })^M \oint \frac{ d z_1}{  z_1} \ldots \oint
\frac{ d z_{M}}{  z_{M} } \prod_{ i \neq j }^M
 ( 1 - \frac{ z_i }{ z_j} )^2 = N^M \frac{ ( 2 M )!}{ 2^M}  \; .
\label{wili}
\eneq
\noindent
Basic properties of $\Psi_{GS}$ follow in this section,
together with their derivation.

\subsection{Singlet State}

The ground state is a spin singlet.  $| \Psi_{GS} \rangle $ is
annihilated by both $S^z$ and $S^-$.  $S^z|\Psi_{GS}\rangle = 0$
because $|\Psi_{GS} \rangle$ has an equal number of $\uparrow$ and
$\downarrow$ spins, while

      \begin{displaymath}
      [S^{-}\Psi_{GS}](z_{2}, \ldots ,z_{M}) = \sum_{\alpha = 1}^{N}
      \langle z_2 , \ldots , z_{M} | S_\alpha^- | \Psi_{GS} \rangle
      \end{displaymath}

      \begin{equation}
        =
      \lim_{z_1 \rightarrow 0} \;
      \sum_{\ell = 1}^{N-1} \frac{1}{\ell !}\biggl\{
      \sum_{\alpha = 1}^{N} z_{\alpha}^{\ell} \biggr\}
      \frac{\partial^{\ell}}{\partial z_{1}^{\ell}} \Psi_{GS}
      (z_{1}, \ldots ,z_{M})  = 0 ,
      \label{singlet}
      \end{equation}

\noindent
      since  $\sum_{\alpha = 1}^{N} z_{\alpha}^{\ell} = N \; \delta_{\ell 0}
      \pmod{N} \; \; \; .$

\noindent

As $\Psi_{GS}$ is a spin singlet, it takes exactly the same form if expressed
either in terms of the $\uparrow$-spin coordinates $z_1 , \ldots , z_M$, or
of the $\downarrow$-spin coordinates, $\eta_1 , \ldots , \eta_M$, that is:

\beq
\Psi_{GS} ( z_1 , \ldots , z_M ) = \Psi_{GS} ( \eta_1 , \ldots , \eta_M ) \;\;\; .
\label{singly} 
\eneq
\noindent
Eq.(\ref{singly}) is proved in Appendix A, where we derive
the formulas to relate the representation of the states of the system
in terms of $\uparrow$-spin coordinates to the representation in terms
of $\downarrow$-spin coordinates.

In the thermodynamic limit, half-odd spin chains exhibit a gapless spectrum,
although they are not allowed to order \cite{conjec}. Accordingly,
$| \Psi_{GS} \rangle$ is a disordered spin liquid state, and the
spin-spin correlation function, $\chi (z_\alpha) =
\langle \Psi_{GS} | S_0^+ S_\alpha^- | \Psi_{GS} \rangle /
\langle \Psi_{GS} | \Psi_{GS} \rangle$, falls off with the distance as
$(-1)^x /x$, thus showing absence of spin order \cite{us2}.

\subsection{Ground State Energy}

At filling-1/2, $| \Psi_{GS} \rangle$ is the ground state of
${\cal H}_{KY}$, with eigenvalue:

\beq
{\cal H}_{KY} | \Psi_{GS} \rangle  = {\cal H}_{HS} | \Psi_{GS} \rangle
= -J \left( \frac{\pi^2}{24} \right)
\left(N + \frac{5}{N} \right) | \Psi_{GS} \rangle  \; \; \; .
\label{gsen}
\eneq
\noindent

Eq.(\ref{gsen}) has been derived originally by Haldane and Shastry
\cite{haldane,shastry}.
In \cite{us2}, we re-derived it by means of our own technique,
consisting in substituting  sums over spins on the lattice with
derivative operators acting on the
analytic extension of $\Psi_{GS} ( z_1 , \ldots , z_{M} )$. 
The $z_j$'s are allowed to take any value on the unit circle. After
computing the derivatives, we constrain them again to lattice
sites. In this subsection we review our technique, in view
of its generalization to the case where the filling is $\neq 1/2$ and
the dynamics of the system is described by the full KY-Hamiltonian.

Since $[S_\alpha^+ S_\beta^- \Psi_{GS} ]
(z_1 , \ldots , z_{M})$ is identically zero unless one of the arguments
$z_1 , \ldots , z_{M}$ equals $z_\alpha$, we have

\begin{displaymath}
[ \biggl\{ \sum_{\beta \neq \alpha}^{N} \frac{ S_{\alpha}^{+}
S_{\beta}^{-}}{\mid \! z_{\alpha} - z_{\beta} \! \mid^2 }
\biggr\} \Psi_{GS}] (z_{1}, \ldots , z_{M})
\end{displaymath}

\begin{displaymath}
= \sum_{j=1}^{M} \sum_{\beta \neq j}^{N} \frac{1}{\mid \! z_j
- z_{\beta} \! \mid^2 } \Psi_{GS} (z_{1}, \ldots , z_{j-1}, z_{\beta} ,
z_{j+1}, \ldots , z_{M})
\end{displaymath}

\begin{equation}
=  \sum_{\ell=0}^{ N -2} \sum_{j = 1}^M \frac{ z_j^{ \ell +1}}{\ell !} A_l
( \frac{ \partial^\ell}{ \partial z_j^\ell} ) \biggl\{
\frac{ \Psi_{GS} ( z_1 , \ldots , z_{M} )}{  z_j} \biggr\} \;\;\; .
\label{engs1}
\end{equation}

\noindent
The coefficients $A_l$ are calculated in \cite{us2}. They are zero for $N>l>2$.
Therefore, Eq.(\ref{engs1}) can be rewritten as:

\begin{displaymath}
 \sum_{j=1}^{M} \biggl\{ \frac{(N-1)(N-5)}{12} z_{j} - \frac{N-3}{2}
 z_{j}^2 \frac{\partial}{\partial z_{j}}
\end{displaymath}

\begin{displaymath}
+ \frac{1}{2} z_{j}^3
 \frac{\partial^2} {\partial z_{j}^2} \biggr\}
  \biggl\{ \frac{ \Psi_{GS} (z_{1},
 \ldots , z_{M})}{ z_{j}} \biggr\}
\end{displaymath}

\begin{displaymath}
= \biggl\{ \frac{N(N-1)(N-5)}{24} - \frac{N-3}{2}
\sum_{j \neq k}^{M} \frac{2 z_{j}}{z_{j} - z_{k}}
\end{displaymath}

\begin{displaymath}
+ \sum_{j \neq k \neq m}^{M} \frac{2 z_{j}^2}
{(z_{j} - z_{k})(z_{j} - z_{m})}+ \sum_{j \neq k}^{M} \frac{ z_{j}^2}
{(z_{j} - z_{k})^2} \biggr\}
\end{displaymath}

\begin{displaymath}
\times  \Psi_{GS} (z_{1}, \ldots ,z_{M})
\end{displaymath}

\begin{equation}
=
\biggl\{ - \frac{N}{8} - \sum_{j \neq k}^{M}
\frac{1}{\mid \! z_{j} - z_{k} \! \mid^2 } \biggr\}
\Psi_{GS} (z_{1}, \ldots , z_{M}) \; \; \; .
\label{kings}
\end{equation}

\noindent
The ``Ising spin term'', on the other hand, provides:

\begin{displaymath}
[ \biggl\{ \sum_{\beta \neq \alpha}^{N} \frac{ S_{\alpha}^{z}
S_{\beta}^{z}}{\mid \! z_{\alpha} - z_{\beta} \! \mid^2 }
\biggr\} \Psi_{GS}] (z_{1}, \ldots , z_{M})
\end{displaymath}

\beq
= \biggl\{ - \frac{N(N^2 - 1)}{48}
+ \sum_{j \neq k}^{M} \frac{1}{\mid \! z_{j} - z_{k} \!
\mid^2 } \biggr\}
\Psi_{GS} (z_{1}, \ldots , z_{M})
\; \; \; .
\label{bishops}
\eneq
\noindent
Adding up Eq.(\ref{kings}) and Eq.(\ref{bishops}) provides Eq.(\ref{gsen}).

Using the factorization property of ${\cal H}_{HS}$, Shastry proved
that $ |\Psi_{GS} \rangle$ is the actual ground state of ${\cal H}_{KY}$
at 1/2-filling \cite{shastry1}. The same proof can be
rephrased within our formalism, as discussed in Ref.\cite{us2}.

The crystal momentum of the state, $q$, is defined (mod $2 \pi$) by the
equation:

\begin{equation}
\Psi_{GS} ( z_1 z , \ldots , z_{M} z ) = e^{ iq}
\Psi_{GS} ( z_1 , \ldots , z_{M}  ) \;\;\;  ,
\label{cm}
\end{equation}
\noindent
where $z$=$\exp (2 \pi i  /N)$ . From Eq.(\ref{cm})
$q$ can be either 0 or $\pi$, according to whether $N$ is divisible by four
or not. In the former case $\Psi_{GS}$ equals itself,  when translated
by one lattice constant, while it equals minus itself in the latter case.

Other relevant properties of $| \Psi_{GS} \rangle $ are discussed in detail in
\cite{us2} and will not be analyzed here.

\section{One-spinon wavefunction.}

The elementary excitations above
the ground state of a correlated 1D electron system are not Landau's
quasiparticles, but rather spinons and holons. Spinons
have been identified as the elementary excitations of a spin-1/2 1D
antiferromagnet. They can be thought of as localized spin defects carrying
total spin-1/2, embedded in an otherwise featureless disordered singlet
spin sea. As spinon excitations appear in  the KY-chain at any filling, we
 study spinons at filling-1/2, when the KY-model reduces back to the
HS-model. One-spinon states appear as states of the chain with an odd
number of sites. In this case, the minimum possible value for the total spin
is 1/2, and the state for a localized spinon at $s$ is created  by
constraining the spin at $s$ to be $\downarrow$, in a surrounding
spin singlet sea \cite{haldane,shastry,us2}. Since correlations in the
ground state are
short-ranged, in the  thermodynamic limit it makes no difference whether one
begins with an odd or an even number of sites. Therefore, in the
thermodynamic limit, there is no way to distinguish between chains with odd
number of sites or chains with even number of sites.
States with half-odd spin are alleged
eigenstates of ${\cal H}_{KY}$ at half-filling with an odd number of spinons
and no holons. In this section we briefly review the one-spinon
wavefunction  and discuss its properties. This was first studied
 by Haldane and Shastry \cite{haldane,shastry},
and discussed at length in \cite{us2} within the
framework of the formalism of analytic variables.

\subsection{One-spinon spin doublet}

Let $N$ be odd and $M=(N-1)/2$. The wavefunction for a localized spinon at
$s$ takes Haldane's form \cite{haldane,shastry}:

\beq
\Psi^{\rm sp}_s( z_1, \ldots , z_M ) = \prod_{j=1}^M ( z_j - s )
z_j \prod_{i < j}^M ( z_i - z_j )^2 \;\;\; ,
\label{locsp}
\eneq
\noindent
where $z_1 , \ldots , z_M$ denote again the position of the
$\uparrow$-spins and $s$ is the coordinate of a lattice site where the spin is
fixed to be $\downarrow$.

By definition, $\Psi^{\rm sp}_s$ is an eigenstate of $S^z$ with eigenvalue
-1/2. In order to prove that it is a spin-1/2 state, we need to show
that $S^-$ annihilates it. Indeed, per Eq.(\ref{singlet}) we have:

\begin{equation}
\sum_{z_\beta \neq z_s}^N S_{\beta}^{-} \Psi^{\rm sp}_{s} = 0
\; \; \; ,
\label{singlet2}
\end{equation}
\noindent
which proves that $\Psi_s^{\rm sp}$ is the $\downarrow$-spin component of a
spin doublet.

$\Psi_s^{\rm sp} (z_1 , \ldots , z_M )$ is a polynomial of
degree less than $N+1$ in each variables $z_j$.  Therefore, we may
again apply Taylor's expansion technique used to calculate the ground
state energy. Doing so, we find:

\begin{displaymath}
{\cal H}_{KY} \Psi_s^{\rm sp} = \frac{J}{2} (\frac{2\pi}{N})^2
\biggl\{ \lambda +\frac{N}{48}(N^2 - 1)
\end{displaymath}

\begin{equation}
 + \frac{M}{6}
(4 M^2 - 1) - \frac{N}{2} M^2 \biggr\} \; \Psi_s^{\rm sp} \; \;\; ,
\label{diffs}
\end{equation}
\noindent
provided that $\lambda$ satisfies the following eigenvalue equation for
$\Phi_s^{\rm sp} = \prod_j^M (z_j - s)$:

\begin{displaymath}
\biggl\{ M ( M - 1) - s^2 \frac{ \partial^2}{ \partial s^2}
- \frac{ N - 3}{2} \left[ M - s \frac{ \partial}{ \partial s }
\right] \biggr\} \Phi_s^{\rm sp}
\end{displaymath}

\begin{equation}
= \lambda \Phi_s^{\rm sp} \;\;\; .
\label{sdif2}
\end{equation}
\noindent
One-$\downarrow$-spinon energy eigenstates are given by propagating
one-spinon plane waves:

\begin{equation}
\Psi_m^{\rm sp} ( z_1 , \ldots , z_M ) = \frac{1}{N} \sum_{s}
 ({s}^{*})^m \Psi_s^{\rm sp} ( z_1 , \ldots , z_M ) .
\end{equation}
\noindent
The energy eigenvalue is

\[
{\cal H}_{KY} | \Psi_{m}^{\rm sp} \rangle
= \biggl\{ - J (\frac{\pi^2}{24}) ( N -
\frac{1}{N})
\]

\beq
+ \frac{J}{2} (\frac{2\pi}{N})^2 m(\frac{N-1}{2} - m)
\biggr\} | \Psi_{m}^{\rm sp} \rangle \;\;\; ,
\end{equation}

\noindent
with $ 0 \leq m \leq (N-1)/2$ and $\lambda = m ( ( N - 1 ) /2 - m)$.

As the total crystal momentum of the state $| \Psi_m^{\rm sp} \rangle $
is given by

\begin{equation}
q_m^{\rm sp} = \frac{\pi}{2} N - \frac{2\pi}{N}(m + \frac{1}{4}) \pmod{2\pi}
\; \; \; ,
\label{momentum}
\end{equation}
\noindent
the total energy may be rewritten as:

\begin{equation}
{\cal H}_{KY} | \Psi_{m}^{\rm sp} \rangle = \biggl\{
- J (\frac{\pi^2}{24}) ( N +
\frac{5}{N} - \frac{3}{N^2}) + E(q_m^{\rm sp}) \biggr\} |
\Psi_{m}^{\rm sp} \rangle
\; \; ,
\end{equation}
\noindent
that is, the sum of a ground-state contribution, plus the kinetic energy of
the propagating spinon. The one-spinon dispersion relation is correspondingly
provided  by:

\begin{equation}
E ( q_m^{\rm sp}) = \frac{J}{2} \biggl[ (\frac{\pi}{2})^2 - (q_m^{\rm sp})^2
\biggr] \pmod{\pi}
\label{dispersion} \;\;\; ,
\end{equation}

\noindent
As extensively discussed in \cite{haldane,shastry,us2}, the one-spinon
dispersion relation shows the typical features of spinon excitations.
It  spans only the inner or outer half of the Brillouin
zone, depending on whether $N - 1$ is divisible
by 4 or not, which corresponds to the absence of negative energy states,
i.e., to the absence of ``antispinons''.
The spinon dispersion at low energies is linear in $q$ with a
velocity

\begin{equation}
v_{\rm spinon} = \frac{\pi}{2} J \; \; \; ,
\label{1dpen}
\end{equation}

\noindent
The half-band of single elementary excitations for odd $N$ are the only
$S=1/2$ states without extra degeneracies. The ground state of the
odd-N spin chain is 4-fold degenerate and is given by $| \Psi_m^{\rm sp}
\rangle$ for $m = 0$ and $(N-1)/2$ and their $\uparrow$
counterparts.  This corresponds physically to a ``left-over'' spinon
with momentum $\pm \pi$.

The spin density in the state $\Psi_s^{\rm sp}$ as a function of
the spinon position is uniformly zero, as appropriate for the
disordered spin singlet, except for an abrupt dip centered at $z=s$
\cite{us2}. The dip is identified with a localized spinon at $s$.
Therefore, $\Psi_s^{\rm sp}$ may be thought of as the wavefunction
for a localized spinon at $s$. 
Starting from such an interpretation, in Ref.(\cite{us2})
 we showed that, although spinons are collective
excitations of a strongly correlated system, they can still be treated
as real quantum mechanical particles. In this paper we will generalize
our formalism to states where both spinons and holons are present.

\subsection{The Norm}

The squared norm of the one-spinon energy eigenstates is defined
as the scalar product:

\beq
\langle \Psi_m^{\rm sp}|\Psi_m^{\rm sp} \rangle= \sum_{ z_1 , \ldots , z_M} |
\Psi_m^{\rm sp} ( z_1 , \ldots , z_M ) |^2  \;\;\; .
\label{nor1}
\eneq
\noindent
For the Haldane-Shastry model, we derived the formula for Eq.(\ref{nor1})
in Ref.\cite{us2}. By employing a recursion relation between $\langle
\Psi_m^{\rm sp}|\Psi_m^{\rm sp} \rangle$ and  $\langle
\Psi_{m-1}^{\rm sp}|\Psi_{m-1}^{\rm sp} \rangle$, we expressed all the norms
in terms of $m$ and of the constant  $C_M$ introduced in Eq.(\ref{wili}).
The induction relation is:

\beq
\frac{\langle \Psi_m^{\rm sp}|\Psi_m^{\rm sp} \rangle}{\langle
\Psi_{m-1}^{\rm sp}|\Psi_{m-1}^{\rm sp} \rangle} =
\frac{ ( m - \frac{1}{2} ) ( M - m + 1 )}{ m ( M - m + \frac{1}{2} )} \;\;\; ,
\label{norma1}
\eneq
\noindent
That recursively gives:

\beq
 \langle \Psi_m^{\rm sp}|\Psi_m^{\rm sp} \rangle  =
\frac{ \Gamma [ M + 1 ] \Gamma [ m + \frac{1}{2} ] \Gamma [ M - m + \frac{1}{2
} ] }{ \Gamma [ \frac{1}{2} ]
\Gamma [ M + \frac{1}{2} ] \Gamma [ m + 1 ] \Gamma [ M - m + 1 ]
} C_M \;\;\; .
\label{norma2}
\eneq
\noindent

\section{One-holon wavefunction.}

Holons are charged, spin-0 elementary excitations of the
Kuramoto-Yokoyama Hamiltonian. They are constructed by removing an
electron from the center of a spinon. The state for a localized holon
at $h_0$ is given by $c_{h_0 \downarrow} | \Psi_{s=h_0}^{\rm sp} \rangle$, 
where $| \Psi_s^{\rm sp} \rangle$ is the state defined in Eq.(\ref{locsp}).
By construction, in the KY-model, the holon is the supersymmetric partner of the
spinon. However, unlike in the spinon case, the Brillouin zone for one-holon
states is not halved, as both negative and positive energy holons can be
constructed. In this section, we are concerned mainly with holon
``kinematics''. We derive the one-holon eigenstates,
their norm, their energy and their crystal momentum. In particular, we
focus on negative-energy one-holon states, since these states are the ones
relevant to the spinon-holon interaction.  In the next Sections we analyze holon 
and spinon dynamics - the interaction
between spinons and holons and its relation to the instability of the
hole excitations in the KY-model.

\subsection{One-Holon Spin Singlet}

Let $N$ be odd and $M = (N-1)/2$. The wavefunction for a propagating,
negative energy, holon is given by:

\beq
\Psi_n^{\rm ho} (z_1 , \ldots , z_M | h ) = (h)^n \prod_j^M ( z_j - h ) z_j
\prod_{j < k}^M (z_j - z_k )^2
\; \; \; ,
\label{holon}
\eneq
\noindent
where $z_1 , \ldots , z_M$ denote the positions of the $\uparrow$ sites
and $h$ denotes the position of the empty site, all others being
$\downarrow$. Also, $ 0 \le n \le (N+1)/2 $. Different from the spinon case,
in Eq.(\ref{holon}) the holon coordinate $h$ is not a quantum number but a 
coordinate variable. Therefore, unlike localized 
one-spinon eigenstates, $\Psi_n^{\rm ho}$
takes a well-defined crystal momentum, as we will show later.

$\Psi_n^{\rm ho}$  is a spin-singlet state. Indeed, by definition its total
component of the spin along $z$ is zero. Following the same
steps leading to Eq.(\ref{singlet2}), we also get:

\beq
S^z  \Psi_n^{\rm ho} = S^-  \Psi_n^{\rm ho} = 0 \; \; ,
\label{singho}
\eneq
\noindent
which proves that $\Psi_n^{\rm ho}$ is a spin singlet.

\begin{figure}
\includegraphics*[width=0.87\linewidth]{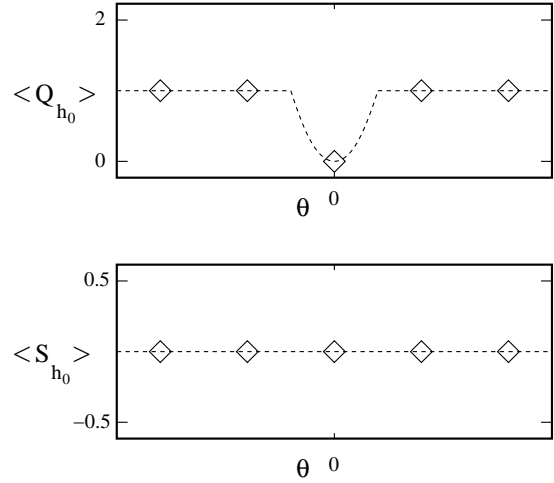}
\caption{Spin and charge profiles of the localized holon
$|\Psi_{h_0}^{\rm ho} \rangle$ defined by Eq.(\ref{localizedholon}). $\theta$
is defined as $\theta = - i \ln (z / h_0 )$, where $z$ is the independent
variable.}
\label{figure1}
\end{figure}

\subsection{Negative One-Holon Energy Eigenstates}

$\Psi_n^{\rm ho}$ is an eigenstate of  ${\cal H}_{KY}$ with energy eigenvalue
given by:

\begin{displaymath}
{\cal H}_{\rm KY} | \Psi_n^{\rm ho} \rangle = \biggl\{ -J
({\frac{\pi^2}{24}})(N-\frac{1}{N})
\end{displaymath}

\begin{equation}
+ \frac{J}{2} (\frac{2\pi}{N})^2 n (n- \frac{N+1}{2}) \biggr\} |
\Psi_n^{\rm ho} \rangle  \;\; ,
\label{negativeholonenergy}
\end{equation}
\noindent
where $0 \le n \le (N+1)/2$.

In order to prove Eq.(\ref{negativeholonenergy}), let us first split
$H_{\rm KY}$ as follows:

\beq
\frac{ H_{\rm KY}}{ \frac{J}{2} ( \frac{2 \pi}{ N } )^2 } =
h_S^T + h_S^V + h_N + h_Q^\downarrow + h_Q^\uparrow \;\; .
\label{split}
\eneq
\noindent
We define the various terms when calculating their contributions to
the total energy.

\begin{itemize}

\item Spin-exchange term:

\[
[ h_S^T \Psi_n^{\rm ho} ] ( z_1 , \ldots , z_M | h )
\]

\[
= [ \sum_{\alpha \neq \beta} \frac{ P S_\alpha^+ S_\beta^- P}{ | z_\alpha -
z_\beta |^2}  \Psi_n^{\rm ho} ] ( z_1 , \ldots , z_M | h )
\]

\[
= \biggl\{ \left[ \frac{1-N^2}{24} - \sum_{ i \neq j}^M \frac{1}{
| z_i - z_j |^2} \right] h^n + \frac{ N-3}{2} h^{n+1} \frac{\partial}{
\partial h}
\]

\beq
 - h^{n+2} \frac{ \partial^2}{ \partial h^2}
\biggr\} \biggl\{ \frac{ \Psi_n^{\rm ho} ( z_1 , \ldots , z_M | h ) }{ h^n}
\biggr\} \; .
\label{piece1}
\eneq
\noindent

\item Spin-Ising term:

\[
[ h_S^V \Psi_n^{\rm ho} ] ( z_1 , \ldots , z_M | h )
\]

\[
= [ \sum_{\alpha \neq \beta} \frac{ P S_\alpha^z S_\beta^z P }{ | z_\alpha -
z_\beta |^2}  \Psi_n^{\rm ho} ] ( z_1 , \ldots , z_M | h )
\]

\[
= \biggl\{ \sum_{i \neq j}^M \frac{1}{ | z_i - z_j |^2} +
\sum_j^M \frac{1}{ | z_j - h |^2}
\]

\beq
- \frac{N (N^2 -1)}{48} \biggr\}
\Psi_n^{\rm ho}  ( z_1 , \ldots , z_M | h ) \;\; .
\label{piece2}
\eneq
\noindent

\item Electrostatic repulsion energy:

\[
[ h_N \Psi_n^{\rm ho} ] ( z_1 , \ldots , z_M | h )
\]

\[
= [ \sum_{\alpha \neq \beta} \frac{ 1}{ | z_\alpha -
z_\beta |^2}  P \biggl[ \frac{1}{2} ( n_\alpha + n_\beta )
\]

\[
 - \frac{1}{4} n_\alpha n_\beta - \frac{3}{4} \biggr] P
\Psi_n^{\rm ho} ]  ( z_1 , \ldots , z_M | h )
\]

\beq
= \frac{1 - N^2}{24} \Psi_n^{\rm ho}  ( z_1 , \ldots , z_M | h ) \;\; .
\label{piece3}
\eneq
\noindent

\item $\downarrow$-charge kinetic energy:

\[
[ h_Q^\downarrow \Psi_n^{\rm ho} ] ( z_1 , \ldots , z_M | h )
\]

\[
= \sum_{\alpha \neq \beta} \left[ \frac{ P c_{\alpha \downarrow} c_{\beta
\downarrow}^\dagger }{ | z_\alpha - z_\beta |^2}
 \Psi_n^{\rm ho} \right]  ( z_1 , \ldots , z_M | h )
\]

\[
= \sum_{z_\beta \neq h} \sum_{k=0}^{M+1} \frac{ z_\beta^n ( z_\beta - h )^k}{
k ! | z_\beta - h |^2} ( \frac{ \partial}{ \partial h} )^k \biggl\{
\frac{ \Psi_n^{\rm ho}   ( z_1 , \ldots , z_M | h ) }{h^n} \biggr\}
\]

\[
= \biggl\{ \left[ \frac{N^2-1}{12} + \frac{n (n-N)}{2} \right] h^n
- \left[ \frac{N-1}{2} - n \right] h^{n+1} \frac{\partial}{\partial h}
\]

\beq
+ \frac{1}{2} h^{n+2} \frac{ \partial^2}{\partial h^2} \biggr\}
\biggl\{ \frac{ \Psi_n^{\rm ho}   ( z_1 , \ldots , z_M | h ) }{h^n} \biggr\} \; .
\label{piece4}
\eneq

\item $\uparrow$-charge kinetic energy:

\[
[ h_Q^\uparrow \Psi_n^{\rm ho} ] ( z_1 , \ldots , z_M | h )
\]

\beq
= \sum_{\alpha \neq \beta} \left[ \frac{ P c_{\alpha \uparrow} c_{\beta
\uparrow}^\dagger }{ | z_\alpha - z_\beta |^2}
 \Psi_n^{\rm ho} \right]  ( z_1 , \ldots , z_M | h ) \; .
\label{piece51}
\eneq
\noindent
To properly work out the contribution in Eq.(\ref{piece51}), we
have to express $\Psi_n^{\rm ho}$ in terms of the $\downarrow$-spin
coordinates, $\eta_1 , \ldots, \eta_M$. In Appendix A we prove that:

\beq
\Psi_n^{\rm ho} (z_1 , \ldots , z_M | h ) = \Psi_n^{\rm ho}
( \eta_1 , \ldots , \eta_M | h ) \; .
\label{piece52}
\eneq
\noindent
Therefore, we obtain:

\[
[ h_Q^\uparrow \Psi_n^{\rm ho} ] ( z_1 , \ldots , z_M | h )
\]

\[
= \biggl\{ \left[ \frac{N^2-1}{12} + \frac{n (n-N)}{2} \right] h^n
- \left[ \frac{N-1}{2} - n \right] h^{n+1} \frac{\partial}{\partial h}
\]

\beq
+ \frac{1}{2} h^{n+2} \frac{ \partial^2}{\partial h^2} \biggr\}
\biggl\{ \frac{ \Psi_n^{\rm ho}   ( \eta_1 , \ldots , \eta_M | h ) }{h^n}
\biggr\} \; .
\label{piece54}
\eneq
\noindent
By using the identity:

\[
h^{n+1} \frac{\partial}{\partial h} \biggl\{ \frac{ \Psi_n^{\rm ho}
( \eta_1 , \ldots , \eta_M | h ) }{h^n} \biggr\}
\]

\[
= \biggl\{ \frac{N-1}{2} - \sum_j^M \frac{h}{h-z_j} \biggr\}
\Psi_n^{\rm ho}   ( z_1 , \ldots , z_M | h )
\]

\beq
= \biggl\{ \left( \frac{N-1}{2} \right) h^n - h^{n+1} \frac{\partial}{\partial
h} \biggr\}  \biggl\{ \frac{ \Psi_n^{\rm ho}
( z_1 , \ldots , z_M | h ) }{h^n} \biggr\} \;\; ,
\label{piece55}
\eneq
\noindent
and the identity:

\[
h^{n+2} \frac{ \partial}{\partial h^2}  \biggl\{ \frac{ \Psi_n^{\rm ho}
( \eta_1 , \ldots , \eta_M | h ) }{h^n} \biggr\}
\]

\[
= \biggl\{ \left[ \frac{(N-1)(N-2)}{3}  + 2 \sum_j^M \left( \frac{h}{ h-z_j}
\right)^2 \right] h^n
\]

\beq
- (N-1) h^{n+1} \frac{ \partial}{ \partial h} + h^{n+2} \frac{ \partial^2}{
\partial h^2} \biggr\}  \biggl\{ \frac{ \Psi_n^{\rm ho}
( z_1 , \ldots , z_M | h ) }{h^n} \biggr\} \;\; ,
\label{piece56}
\eneq
\noindent
we finally derive:

\[
[ h_Q^\uparrow \Psi_n^{\rm ho} ] ( z_1 , \ldots , z_M | h ) =
\]

\[
\biggl\{ \frac{n (n-1)}{2} h^n + \sum_j^M \frac{h^{n+2}}{ (h - z_j )^2}
- n h^{n +1} \frac{ \partial}{ \partial h}
\]

\beq
+ \frac{1}{2} h^{n+2} \frac{ \partial^2}{ \partial h^2} \biggr\}
\Psi_n^{\rm ho}  ( z_1 , \ldots , z_M | h ) \;\; .
\label{piece57}
\eneq

\end{itemize}

Adding Eqs.(\ref{piece1},\ref{piece2},\ref{piece3},\ref{piece4},\ref{piece57})
all together, we find that:

\[
[ H_{\rm KY} \Psi_n^{\rm ho} ] ( z_1 , \ldots , z_M | h )
\]

\[
= \frac{J}{2} \left( \frac{ 2 \pi}{N} \right)^2
\biggl\{ - \frac{N (N^2 -1 )}{48}
\]

\beq
+ n ( n - M  - 1 ) \biggr\} \Psi_n^{\rm ho}  ( z_1 , \ldots , z_M | h )
\;\; ,
\label{enerproff}
\eneq
\noindent
This is the formula we had to prove.

\subsection{Crystal Momentum}

The state $|\Psi_n^{\rm ho} \rangle $ is a propagating holon with crystal
momentum

\begin{equation}
q_n^{\rm ho} = \frac{\pi}{2} N + \frac{2\pi}{N} (n- \frac{1}{4}) \;\;
\pmod{2\pi} \;\; ,
\label{holoncrystalmomentum}
\end{equation}
\noindent
with the definition
\begin{equation}
\Psi_n^{\rm ho}(z_1 z,\ldots,z_M z|hz)=\exp(iq_n^{\rm ho})
\Psi_n^{\rm ho}(z_1 ,\ldots,z_M |h)
\;\;\; .
\label{scalingholon}
\end{equation}
\noindent
Rewriting the eigenvalue as

\begin{equation}
{\cal H}_{KY} | \Psi_{n}^{\rm ho} \rangle = \biggl\{ -
J (\frac{\pi^2}{24}) ( N + \frac{5}{N} - \frac{3}{N^2}) + E(q_n^{\rm ho})
\biggr\} | \Psi_{n}^{\rm ho} \rangle
\; \; ,
\end{equation}
\noindent
we obtain the dispersion relation

\begin{equation}
E ( q_n^{\rm ho}) = - \frac{J}{2} \biggl[ (\frac{\pi}{2})^2 - (q_n^{\rm ho})^2
\biggr] \pmod{\pi}
\label{negativedispersion} \;\;\; ,
\end{equation}
\noindent
We worked out Eq.(\ref{negativedispersion}) for the case of
negative-energy one-holon eigenstates. Unlike in the spinon case, the holon
dispersion relation around the band minimum is quadratic in $q$, while it is
linear in $q$ near $\pi/2$, where the band closes.

Positive-energy one-holon eigenstates
may be constructed by supersimmetrically rotating  one-spinon eigenstates
\cite{chia}. Their energy is given by $ | E ( q_n^{\rm ho} ) |$ and the
momentum spans the remaining half of the Brillouin zone. The  corresponding
dispersion  relation is plotted in Fig.(\ref{figure1}). Since for the
purpose of studying spinon-holon interaction we only need negative-energy
holon states only, we do not discuss here positive-energy holon
states. We also need the wavefunction for a localized holon at site
$h_0$, $|\Psi_{h_0}^{\rm ho} \rangle$, which is obtained by Fourier
transforming the propagating holon wavefunction to real space:
\begin{equation}
\Psi_{h_0}^{\rm ho} = \sum_{n=0}^{(N+1)/2} h_0^{-n} \Psi_n^{\rm ho}
\label{localizedholon}
\end{equation}

\begin{figure}
\includegraphics*[width=0.87\linewidth]{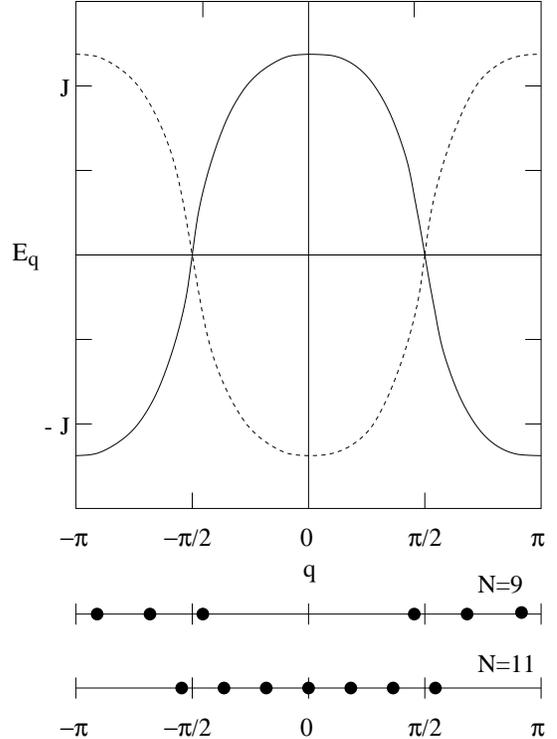}
\caption{Upper graph: Holon dispersion relation. Positive energy holons are unstable towards 
decay into spinons; Lower: Configuration space for N=9 and N=11 sites}
\label{figure2}
\end{figure}

\subsection{The Norm}

The norm of the state $\Psi_n^{\rm ho}$ is defined as:

\beq
\langle \Psi_n^{\rm ho} | \Psi_n^{\rm ho} \rangle
= \sum_{z_1 , \ldots , z_M;  h} | \Psi_n ( z_1 , \ldots , z_M | h ) |^2
\label{oneholonor}
\eneq
\noindent
From the definition of the norm it immediately follows out that $\langle
\Psi_n^{\rm ho} | \Psi_n^{\rm ho} \rangle$ does not depend on $n$. The
formula for $\langle \Psi_n^{\rm ho} | \Psi_n^{\rm ho} \rangle$ is
worked out in Appendix B and is given by:

\beq
\langle \Psi_n^{\rm ho} | \Psi_n^{\rm ho} \rangle =
N^{M+1} \frac{ (2M)!}{ 2^M} \frac{ \Gamma [ \frac{1}{2} ] \Gamma [
M + 1 ]}{\Gamma [ M + \frac{1}{2} ] } \;\; .
\label{oneholonor2}
\eneq

\section{One-spinon One-holon wavefunction}

As for many-spinon configurations, spinons and holons maintain their
integrity when many of them are present in the same state. Therefore, we
can diagonalize ${\cal H}_{KY}$ within subspaces with a fixed number
of spinons and holons.

In this section we derive the action of ${\cal H}_{KY}$ on the one-spinon
one-holon eigenstates, diagonalize the corresponding matrix and work out the
norm of the states.

\subsection{Action of  ${\cal H}_{KY}$ on one-spinon one-holon states}

To construct one-spinon one-holon eigenstates of
$ {\cal H}_{\rm KY} $ we start with states in a mixed representation,
$\Psi_n^s$, where the spinon is localized at site $s$, but the holon is
propagating with momentum $q_n^{\rm ho}$. For $N$ even and $M=N/2 -1$,
we have

\begin{equation}
\Psi^n_s (z_1, \ldots , z_M|h)= h^n  \prod_j^M (z_j - s ) ( z_j - h ) z_j
\prod_{j < k}^M (z_j - z_k )^2
 ,
\label{localizedspinonpropagatingholon}
\end{equation}
\noindent
where $1 \le n \le M+2$. To derive the action of the KY-Hamiltonian on
$\Psi_n^s$, we split it as in Eq.(\ref{split}).

\begin{itemize}

\item The action of spin-exchange term on $\Psi_s^n$ provides:

\[
[ h_S^T  \Psi_s^n ] ( z_1 , \ldots , z_M | h ) =
\]

\[
- \sum_{ j = \uparrow} \sum_{ z_\beta \neq z_j } \frac{ z_j z_\beta^2}{
( z_j - z_\beta )^2} \Psi_s^n ( z_1 , \ldots , z_\beta , \ldots | h )
\]

\[
+ \frac{M}{12} ( -2N + 5) \Psi_s^n  - \sum_{ i \neq j= \uparrow} \frac{1}{
| z_i - z_j|^2} \Psi_s^n
\]

\[
+ (M-1) \left[ s \frac{ \partial }{ \partial s} \Psi_s^n + h^{n+1} \frac{
\partial }{ \partial h} \left( \frac{ \Psi_s^n }{ h^n }  \right)\right]
\]

\[
- s^2 \frac{ \partial^2}{ \partial s^2} \Psi_s^n  - h^{ n+2} \frac{
\partial^2}{ \partial h^2}  \left( \frac{ \Psi_s^n }{ h^n } \right)
\]

\beq
+ \frac{1}{2} \left( \frac{ h +s}{ h - s} \right) \left[ s \frac{ \partial}{
\partial s} \Psi_s^n - h^{n+1} \frac{ \partial}{ \partial h}
\left( \frac{ \Psi_s^n }{ h^n } \right) \right] \; .
\eneq

\item The Ising term gives:

\[
[ h_S^V \Psi_s^n ] (z_1 , \ldots , z_M | h ) =
\]

\beq
\biggl\{ \sum_{ i \neq j=\uparrow} \frac{1}{ | z_i - z_j |^2}
- \frac{M}{2} \frac{ N^2 - 1}{12} + \sum_{ j = \uparrow} \frac{1}{ | z_j -
h |^2} \biggr\} \Psi_s^n \; .
\eneq

\item The Coulomb potential term acting on the one-spinon-one-holon
wavefunction reads:

\beq
[ h_N \Psi_s^n ] (z_1 , \ldots , z_M  | h ) =
\frac{ 1 - N^2}{24} \Psi_s^n \;\; .
\eneq

\item  The $\downarrow$ spin contribution to the charge kinetic energy gives:

\[
[ h_Q^\downarrow \Psi_s^n ]  (z_1 , \ldots , z_M  | h ) =
\]

\[
- \sum_{z_\beta \neq h} \frac{z_\beta h}{ ( z_\beta - h )^2 }
\Psi_s^n  (z_1 , \ldots , z_M  | z_\beta )
\]

\[
= \left[ \frac{ N^2 - 1}{12} + \frac{ n ( n - N )}{2} \right] \Psi_s^n
\]

\beq
- \left[ \frac{ N-1}{2} - n \right] h^{n+1} \frac{ \partial }{ \partial h}
\left( \frac{ \Psi_s^n}{ h^n} \right)
+
\frac{1}{2} h^{ n+2} \frac{ \partial^2}{ \partial h^2}
 \left( \frac{ \Psi_s^n}{ h^n} \right) .
\eneq

\item In order to manipulate the $\uparrow$-spin contribution to the charge
kinetic energy, we need to express $\Psi_s^n ( z_1 , \ldots , z_M | h)$ in
terms of the $\downarrow$ spin coordinates $\eta$. As in the one-holon case,
we obtain:

\[
\Psi_s^n ( z_1 , \ldots , z_M | h )=
\Psi_s^n ( \eta_1 , \ldots , \eta_M | h ) \;\;\; .
\]
\noindent
Therefore, we have:

\[
[ h_Q^\uparrow \Psi_s^n ] ( \eta_1 , \ldots , \eta_M | h ) =
\]

\[
- \sum_{ z_\beta \neq h} \frac{ h z_\beta}{ ( h - z_\beta )^2}
\Psi_s^n ( \eta_1 , \ldots , \eta_M | z_\beta ) 
\]

\[
+ \frac{ sh}{ ( s-h)^2}\Psi_s^n ( \eta_1 , \ldots , \eta_M | s )
\]

\[
= \left[ \frac{N^2 - 1}{12} + \frac{ n ( n - N )}{2} \right] \Psi_s^n
\]

\[
- \left[ \frac{N-1}{2} - n \right] h^{n+1} \frac{ \partial}{ \partial h}
\left( \frac{\Psi_s^n }{ h^n} \right) 
\]

\[
+ \frac{1}{2} h^{n+2} \frac{ \partial^2
}{ \partial h^2} \left( \frac{\Psi_s^n }{ h^n} \right)
\]

\beq
+
\frac{ sh}{ ( s-h)^2}\Psi_s^n ( \eta_1 , \ldots , \eta_M | s ) \;\;\; .
\label{bour}
\eneq
\noindent
By going back to $\uparrow$-spin coordinates $z$, Eq.(\ref{bour}) provides:

\[
\frac{ n^2 - 2 n + 1}{2} \Psi_s^n + \left( \frac{1-n}{2} \right)
\frac{ h+ s}{h - s} \Psi_s^n
\]

\[
+ \left(\frac{3}{2} -n \right) h^{n+1} \frac{ \partial}{ \partial h}
\left( \frac{ \Psi_s^n }{ h^n} \right) + \frac{ hs}{ (h-s)^2} \Psi_s^n
\]

\[
+\frac{1}{2}  \left( \frac{ h +s}{ h - s} \right) h^{n+1}
\frac{ \partial}{ \partial h} \left( \frac{ \Psi_s^n}{ h^n} \right)
+ \frac{1}{2} h^{n+2} \frac{ \partial^2}{ \partial h^2} \left(
\frac{ \Psi_s^n }{ h^n} \right)
\]

\beq
- \sum_{ z_j} \frac{1}{ | z_j - h |^2}
\Psi_s^n -\frac{sh}{ ( s - h )^2} \left( \frac{s}{h} \right)^{n-1}
\Psi_h^n \; .
\label{reverse}
\eneq
\noindent
\end{itemize}
\noindent
Adding up all the contributions together, we obtain:

\[
h_{\rm KY} \Psi_s^n ( z_1 , \ldots , z_M | h ) =
\]

\[
\biggl\{ \frac{-N^3 + 19 N}{48} + \left( n - \frac{N}{2} - 1 \right)
\biggr\} \Psi_s^n
\]

\[
+ ( M-1) s \frac{ \partial}{ \partial s} \Psi_s^n - s^2 \frac{ \partial^2}{
\partial s^2} \Psi_s^n
\]

\[
+\frac{1}{2} \frac{ h + s}{ h-s} \left( s \frac{ \partial}{ \partial s}
\Psi_s^n + (1-n) \Psi_s^n \right) 
\]

\beq
+ \frac{hs}{ (h-s)^2} \left( \Psi_s^n - \left( \frac{s}{h} \right)^{n-1}
\Psi_h^n \right) \;\;\; .
\label{eqmo} 
\eneq
\noindent
In the next subsection we will will solve Eq.(\ref{eqmo}) by working out the
basis of one-spinon one-holon states that diagonalize $h_{KY}$.

\subsection{One-spinon one-holon energy eigenstates.}

To diagonalize $h_{KY}$, we introduce the propagating one-spinon
one-holon energy eigenstates

\begin{equation}
\Psi_{mn} (z_1, \ldots , z_M|h) = \sum_s \frac{s^{-m}}{N} \Psi^n_s
(z_1 , \ldots , z_M | h ) \; .
\label{propagatingspinonpropagatingholon}
\end{equation}
\noindent
The second and the third rows of Eq.(\ref{eqmo}) are diagonal in the basis
of the states $\Psi_{mn}$, and their contribution is given by:

\beq
[ \epsilon_0 + n ( n - 1 - \frac{N}{2} ) +
m ( M - m ) ] \Psi_{mn} \;\;\; .
\eneq
\noindent
where:

\[
\epsilon_0 = \frac{ - N^3 + 19 N}{48} \;\;\; .
\]
\noindent
On the other hand, the diagonalization of the ``interaction'' 
term, given by the fourth and the
fifth rows of Eq.(\ref{eqmo}), needs further work.
 By using a straightforward, although tedious, application
of basic identities proved in the Appendix of Ref.\cite{us2}, one obtains:

\[
\sum_{s \in S^N } s^{ -m} \biggl\{
\frac{1}{2} \left( \frac{ h + s}{ h - s} \right) \left[ s \frac{ \partial}{
\partial s} \Psi_s^n + ( 1 - n )  \Psi_s^n \right]
\]
\[
+ \frac{ hs}{ ( h - s)^2} \left[  \Psi_s^n - \left( \frac{ s}{ h} \right)^{n-1}
 \Psi_h^n \right] \biggr\} 
\]

\[
= \sum_{s \neq h } s^{ -m} \biggl\{
\frac{1}{2} \left( \frac{ h + s}{ h - s} \right) \left[ s \frac{ \partial}{
\partial s} \Psi_s^n + ( 1 - n )  \Psi_s^n \right]
\]
\[
+ \frac{ hs}{ ( h - s)^2} \left[  \Psi_s^n - \left( \frac{ s}{ h} \right)^{n-1}
 \Psi_h^n \right] \biggr\} +
\]

\[
\lim_{s \rightarrow h} s^{ -m} \biggl\{
\frac{1}{2} \left( \frac{ h + s}{ h - s} \right) \left[ s \frac{ \partial}{
\partial s} \Psi_s^n + ( 1 - n )  \Psi_s^n \right]
\]
\[
+ \frac{ hs}{ ( h - s)^2} \left[  \Psi_s^n - \left( \frac{ s}{ h} \right)^{n-1}
 \Psi_h^n \right] \biggr\}=
\]

\beq
- \frac{N}{2} ( m -n+1) \Psi_{mn} + \sum_{ j =0}^m N (m-n+1 ) \Psi_{ m-j,n-j} \; ,
\eneq
\noindent
if $n-m-1 > 0$ and

\beq
- \frac{N}{2} ( n-m-1) \Psi_{mn} + \sum_{j=0}^{M-m} N (n-m - 1) \Psi_{m+j,n+j} \; ,
\eneq
\noindent
if $n-m-1 \leq 0$. Therefore the action of $h_{\rm KY}$ on
$\Psi_{mn}$ is given by:

\[
[ h_{\rm KY} \Psi_{mn} ] =
\left[ \epsilon_0 + m ( M - m ) + n ( n - 1  - \frac{N}{2} ) \right]
\Psi_{mn}
\]

\beq
- \frac{1}{2} ( n - m - 1 ) \Psi_{mn} - ( n - m - 1 ) \sum_{j=1}^m
\Psi_{m-j, n-j} ,
\eneq

\noindent
if $m-n+1 < 0$, and

\[
[ h_{\rm KY} \Psi_{mn} ] =
\left[ \epsilon_0 + m ( M - m ) + n ( n - 1  - \frac{N}{2} ) \right]
\Psi_{mn}
\]

\beq
- \frac{1}{2} ( m - n + 1 ) \Psi_{mn}
-  ( m - n + 1 ) \sum_{j=1}^{M-m} \Psi_{m+j, n+j} , 
\eneq
\noindent
if  $m-n+1 \geq 0$.
Since the Hamiltonian matrix is upper (or lower) triangular, complete
diagonalization is possible. The energy eigenstates will be linear
combinations of $\Psi_{mn}$

\beq
\Phi_{mn} = \sum_{j=0}^m a_j \Psi_{ m - j , n - j} \;\;\; ,
\label{linearcombinationspinonholon1}
\eneq
\noindent
if $m-n+1 < 0$, and:

\beq
\Phi_{mn} = \sum_{j=0}^{M-m} a_j \Psi_{ m + j , n + j} \;\;\; ,
\label{linearcombinationspinonholon2}
\eneq
\noindent
if $m-n+1 \geq 0$, corresponding to the energy eigenvalues

\[
E_{mn}^+ = \frac{J}{2} \left( \frac{2 \pi}{N} \right)^2 \biggl[
\epsilon_0 + m ( M - m )
\]

\[
+ n ( n - 1  - \frac{N}{2} )
- \frac{1}{2} ( n - m - 1 ) \biggr] \;\;\; ,
\]
\noindent
if $m-n+1 < 0$, and:

\[
E_{mn}^- =\frac{J}{2} \left( \frac{2 \pi}{N} \right)^2 \biggl[
 \epsilon_0 + m ( M - m )
\]

\[
 + n ( n - 1  - \frac{N}{2} )
+ \frac{1}{2} ( n - m - 1 ) \biggr] \;\;\; , 
\]
\noindent
if $m-n+1 \geq 0$. The coefficients $a_l$ are defined by the recursion
relation:

\beq
a_l=-\frac{1}{2l} \sum_{k=0}^{l-1} a_k \;\;\;\; a_0=1 \;\;\; .
\eneq
\noindent
In terms of spinon and holon momenta, $E^+_{mn}$ and $E^-_{mn}$ take the same
form $E_{mn}$, given by:

\beq
E_{mn}=E_{GS}+E(q_m^{\rm sp})+E(q_n^{\rm ho})-\frac{\pi J}{N}
\frac{|q_m^{\rm sp} - q_n^{\rm ho}|}{2} .
\eneq
\noindent
$E_{mn}$ is the sum of the ground-state energy, the energies of an isolated
spinon and an isolated holon plus a negative interaction contribution that
becomes negligibly small in the thermodynamic limit.
Eqs.(\ref{linearcombinationspinonholon1},
\ref{linearcombinationspinonholon2}) can
be inverted. The result is:

\beq
\Psi_{mn} = \sum_{j=0}^m b_j \Phi_{ m - j , n-j} \;\;\; ,
\label{inverselinearcombinationspinonholon1}
\eneq
\noindent
if $m-n+1 < 0$, and:

\beq
\Psi_{mn} = \sum_{j=0}^{M-m} b_j \Phi_{ m + j , n+j} \;\;\; ,
\label{inverselinearcombinationspinonholon2}
\eneq
\noindent
if $m-n+1 \geq 0$.
The coefficients are given by:

\beq
b_j = \frac{ \Gamma [ j + \frac{1}{2} ]
}{ \Gamma  [ \frac{1}{ 2} ]  \Gamma [ j + 1 ] } \;\;\; .
\label{eq17}
\eneq
\noindent

\subsection{The Norm}

The squared norm of the state $\Psi_{mn}$ is defined as:

\beq
\langle \Phi_{mn}| \Phi_{mn} \rangle =
\sum_{ z_1 , \ldots , z_M } | \Phi_{mn} ( z_1 ,
\ldots , z_M |) |^2 \;\;\; .
\label{eqb1}
\eneq
\noindent
In a similar fashion to the two-spinon case discussed in \cite{us2}, we
compute the norm of the one-spinon one-holon states by means of
mathematical induction. The calculation is presented in detail in
Appendix C. The basic induction relation in the case $n-k+1 < 0$ is given by:

\beq
\frac{ \langle \Phi_{kn} | \Phi_{kn} \rangle}{ \langle \Phi_{k-1,n} |
\Phi_{k-1,n} \rangle} = \frac{ ( k - \frac{1}{2} ) ( M - k + \frac{3}{2} )}{
k (M-k+1)} \;\; ,
\label{eqc18}
\eneq
\noindent
which provides the formula for the norm of the one-spinon one-holon energy
eigenstates:

\beq
\langle \Phi_{kn} | \Phi_{kn} \rangle = N^{M+1} \frac{ (2M)!}{2^M}
( M + \frac{1}{2} ) \frac{ \Gamma [ M - k + 1] \Gamma [ k +
\frac{1}{2}]}{ \Gamma [ M - k + \frac{3}{2} ] \Gamma [ k + 1 ]} .
\label{eqc19}
\eneq
\noindent
In the complementary case, $n-k+1 \geq 0$, we obtain:

\[
\langle \Phi_{kn} | \Phi_{kn} \rangle = \langle \Phi_{M-k,M-n} |
\Phi_{M-k,M-n} \rangle
\]

\beq
=  N^{M+1} \frac{ (2M)!}{2^M} ( M + \frac{1}{2} )
\frac{ \Gamma [ M-k+\frac{1}{2} ] \Gamma [ k + 1 ] }{ \Gamma [ M - k +1 ]
\Gamma [ k + \frac{3}{2} ] } \; .
\label{eqc20}
\eneq

\begin{figure}
\includegraphics*[width=0.97\linewidth]{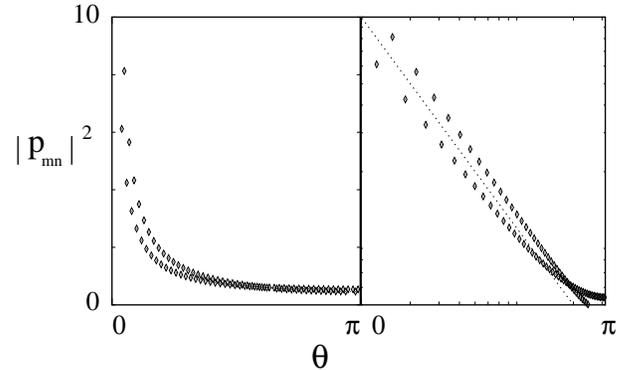}
\caption{Left panel: $|p_{mn} ( e^{i \theta} )|^2$ vs. $\theta$ for $m=M, n=0$;
Right panel: the same plot on a log-log scale. The dashed straight line is a
plot of $1/\theta$}
\label{figure3}
\end{figure}

\section{Spinon-Holon Attraction}

In this Section we analyze the interaction between a spinon and a holon by
constructing the real-space representation of the one-spinon
one-holon wavefunction, and by studying the behavior of the corresponding
probability as a function of the separation between the two particles.
Our exact results show that a spinon and a holon interact through
 a short-range attraction identical, in the thermodynamic limit,
 to the atraction between two spinons.

The state for a localized spinon at site $s$ and a localized holon at site
$h_0$, $\Psi_{sh_0}$ is defined as the Fourier transform of $\Psi_s^n$ back
to coordinate space:

\beq
\Psi_{sh_0}= \sum_{n=1}^{M+2} h_0^{-n}\Psi_s^n \;\;\; .
\label{localizedspinonlocalizedholonwavefunction}
\eneq
\noindent
Following the same steps as for the two-spinon wavefunctions
\cite{us1,us2}, we define the real-space coordinate representation for
a spinon-holon pair, $s^mh_0^{-n}p_{mn}(s/h_0)$, as follows:

\[
\Psi_{s h_0} = \sum_{ n = 1}^{ M + 2 } \sum_{ m = 0 }^{n-2}
s^m h_0^{-n} \sum_{ j = 0}^m b_j \Phi_{ m - j , n - j }
\]

\[
+ \sum_{ n = 1}^{M+2} \sum_{ m = n- 1}^M s^m h_0^{-n}
\sum_{ j = 0}^{M-m} b_j \Phi_{m+j , n + j}
\]

\[
= \sum_{n=1}^{M+2} \sum_{m=0}^{n-2} s^m h_0^{-n} p_{mn} ( \frac{s}{ h_0} )
\Phi_{mn} 
\]

\beq
+ \sum_{ n = 1}^{M+2 } \sum_{ m = n - 1}^M s^m h_0^{-n}
p^{'}_{mn} ( \frac{s}{ h_0} ) \Phi_{mn} \;\;\; ,
\label{firsteqinsection}
\eneq
\noindent
where $|\Phi_{mn} \rangle$ is an eigenstate of ${\cal H}_{KY}$ with
eigenvalue $E_{mn}$, that is:

\beq
\langle\Phi_{mn} | H_{\rm KY} | \Psi_{sh_0} \rangle = E_{mn} \langle
\Phi_{mn} | \Psi_{sh_0} \rangle \;\; .
\label{firsts}
\eneq
\noindent
The matrix element $\langle \Phi_{mn} | {\cal H}_{KY} | \Psi_{s h_0}
\rangle$ can be written as a differential operator acting on the analytic
extension of $\langle \Phi_{mn} | \Psi_{s h_0} \rangle $, where $s$ and $h_0$
are understood to take any value on the unit circle. Therefore, by equating
$\langle \Phi_{mn} | {\cal H}_{KY} | \Psi_{s h_0} \rangle$ to $E_{mn}
\langle \Phi_{mn} | \Psi_{s h_0} \rangle$, it is straightforward to write
down the equation of motion for the one-spinon one-holon wavefunction, which
reads:

\[
( E_{mn} - E_{\rm GS})  \langle \Phi_{mn} |  \Psi_{s h_0} \rangle  =
 \langle \Phi_{mn} | ( H_{\rm KY}- E_{\rm GS})  | \Psi_{s h_0} \rangle =
\]

\[
J ( \frac{ 2 \pi}{N} )^2 \biggl\{ \biggl[   ( M
- s \frac{ \partial}{ \partial s} ) s \frac{ \partial}{ \partial s}
+  h_0 \frac{ \partial }{ \partial h_0} ( 1 + \frac{N}{2} + h_0 \frac{ \partial}{
\partial h_0} )
\]

\[
 + \frac{1}{2} \left( \frac{ h_0 + s}{ h_0 - s} \right) \left(
s\frac{ \partial}{ \partial s} + h_0 \frac{ \partial }{\partial h_0} + 1
\right) \biggr]   \langle \Phi_{mn} |  \Psi_{s h_0} \rangle
\]

\beq
+ \frac{h_0}{s - h_0} \left( \frac{s}{h_0} \right)^\nu
\langle \Phi_{mn} |  \Psi_{h_0 h_0} \rangle
\biggr\} \;\;\; ,
\label{extendeddiffeq}
\eneq
\noindent
where $\nu = M$ if $m-n+1 < 0$, $\nu = 0 $ otherwise. In the differential
operator in Eq.(\ref{extendeddiffeq}),  we recognize the sum of the
energies of the free spinon and holon, a velocity-dependent
interaction, which diverges at small spinon-holon separation,
and another term which takes into account the
correction for the case when the spinon and the holon are at the same
position. By using
Eqs.(\ref{firsteqinsection},\ref{firsts},\ref{extendeddiffeq}),
we find the following equations for the ``relative
wavefunctions'' $p_{mn}(z)$ and $p_{mn}^{'}(z)$ ($z=s /h_0$):

\beq
[ 2 \frac{d}{d z} - \frac{1}{ (1-z)} ] p_{mn} (z) + \frac{z^{M-m-1}}{ (1-z) }
 p_{mn} (1) = 0 \; , 
\label{eq18}
\eneq
\noindent
if $m-n+1 <0$, and

\beq
\left[ 2 \frac{d}{d ( \frac{1}{z} )} -
\frac{1}{ ( 1 - \frac{1}{z} )} \right] p_{mn}^{'} ( z )
+ \frac{ ( \frac{1}{z} )^m }{  ( 1 - \frac{1}{z} ) } p_{mn}^{'} ( 1 ) = 0 \; ,
\label{eq19}
\eneq
\noindent
if  $m-n+1 \geq 0$.

Eqs.(\ref{eq18},\ref{eq19}) are first-order ``Dirac-like'' equations.
They are first order because the spinon and the holon energy bands have
opposite curvature. In this respect, they differ from the differential
equation obtained in the two spinon case, which was second order
\cite{us1,us2}. The corresponding solutions are given by:

\beq
p_{mn} ( z ) = \sum_{k = 0}^{M-m-1} \frac{ \Gamma [ k + \frac{1}{2} ] }{
\Gamma [ \frac{1}{2} ] \Gamma [ k + 1 ] } z^k \;\;\; ,
\label{eq20}
\eneq
\noindent
for Eq.(\ref{eq18}) and

\beq
p_{mn}^{'} ( z ) = \sum_{k = 0}^{m} \frac{ \Gamma [ k + \frac{1}{2} ] }{
\Gamma [ \frac{1}{2} ] \Gamma [ k + 1 ] } (\frac{1}{z})^k \;\;\; .
\label{eq21}
\eneq
\noindent
for Eq.(\ref{eq19}).

 The value of the spinon-holon wavefunction at zero separation between the
particles is derived in Appendix D. It is given by:

\beq
p_{mn}(1)=2 \frac{ \Gamma [ M-m + \frac{1}{2} ]}{
\Gamma [ \frac{1}{2} ] \Gamma [ M-m ] } \;\;\; ,
\label{pmn1}
\eneq
\noindent
and

\beq
p_{mn}^{'}(1)=2 \frac{ \Gamma [m + \frac{3}{2} ]}{
\Gamma [ \frac{1}{2} ] \Gamma [ m+1 ] } \;\;\; ,
\label{pmn2}
\eneq
\noindent
Within the framework of our formalism, it
is possible to treat spinons and holons, collective excitations
of strongly-correlated one-dimensional electron systems, as actual
quantum-mechanical particles. We were first able to associate
 a two-particle wavefunction to a spinon-holon pair and to write down the
corresponding equation of motion
(Eqs.(\ref{firsteqinsection},\ref{extendeddiffeq})). We then worked out
the exact wavefunctions corresponding to each energy eigenvalue
(Eqs.(\ref{eq20},\ref{eq21})).

The squared modulus for the spinon-holon wavefunction, $|p_{mn}(z)|^2$,
gives the probability for a spinon and a holon configuration as a
function of the separation between the two particles. In Fig.\ref{figure3}
we plot  $|p_{mn}( e^{i \theta} )|^2$ versus the distance between the spinon
and the holon, $\theta$. From  Fig.\ref{figure3} the nature of the interaction
between a spinon and a holon may be easily inferred. While at large
separations  $|p_{mn}( e^{i \theta} )|^2$ does not depend on $\theta$, as
it is appropriate for noninteracting particles,  at small separations
it shows a remarkable enhancement. This corresponds to a huge increase of
the probability of configurations with the spinon and the holon on top
of each other. The enhancement does not depend on the holon momentum.
As $N$ gets larger, the probability enhancement
peaks up. It survives the thermodynamic limit,
even though the interaction energy goes to zero, and the total energy becomes
the sum of the energies of the isolated spinon and holon. However, the
attraction is not strong enough to create a spinon-holon bound state, even
in the thermodynamic limit. This corresponds to the absence of a low-energy
stable hole excitation, and it is what causes the quasiparticle peak to
disappear.

An intriguing feature of the spinon-holon interaction in the thermodynamic
limit is that it has the same power-law form as the spinon-spinon
interaction derived in \cite{us1,us2}. In the right panel of Fig.\ref{figure3}
we plot $|p_{mn} ( e^{i \theta} ) |^2$ on a log-log scale and compare it
with $1/\theta$. The probability falls off as the first
power of the separation between the two particles. This shows that, although
the equation of motion for a spinon and a holon is quite different
from the one for two spinons, the interaction in
both cases results in a short-range attraction, and its effects on the
corresponding two-particle wavefunction are basically the same.

\section{Hole Spectral Function}

In this Section we work out $A_{\rm h}^{\rm sp, ho} ( \omega , q)$, the
one-spinon one-holon contribution to the hole spectral function
$ A_{\rm h} (\omega , q)$. We show that this contribution
(which provides quite a good approximation to $ A_{\rm h} (\omega , q)$
for $q \sim 0$) depends only on the $p_{mn}$'s and the  $p_{mn}^{'}$'s
calculated at $z=1$ (that is, as the spinon and the
holon lay at the same site). This allows us to obtain for any finite $N$ a
simple closed-form expression for  $A_{\rm h}^{\rm sp, ho} ( \omega , q)$,
and to relate it to the spinon-holon interaction. In  the thermodynamic
limit, we obtain the previously known formula for the contribution
of the one-spinon one-holon states to  $A_{\rm h}^{\rm sp, ho}
( \omega , q)$ \cite{kato}. The formula in \cite{kato} shows that
there is no low-energy hole pole in the hole spectral function, but rather
a sharp square-root singularity followed by a branch cut. These features
have also been experimentally detcted by means of ARPES experiments on
quasi 1D insulator \cite{zxshen}.

The branch cut corresponds to the lack of integrity of the hole excitation,
which breaks up into a spinon and a holon. Here we will show that, in the
thermodymanic limit, the probability enhancement $p_{mn} (1)$
($p_{mn}^{'} (1)$) turns into the square-root singularity at threshold for
a spinon-holon pair. As a consequence, we prove that the square-root
singularity in the hole spectral function is a direct consequence
of the interaction between spinons and holons. Therefore, it can be
directly experimentally measured.

We begin with the calculation of $A_{\rm h}^{\rm sp , ho} (\omega , q)$
for a finite lattice. In Lehman representation we obtain:

\[
A_{\rm h}^{\rm sp, ho} (\omega , q ) = \Im m  \biggl\{ \sum_X \frac{
| \langle X | \sum_{h_0} (h_0)^{-k}
 c_{h_0 \uparrow} | \Psi_{\rm GS} \rangle |^2 }{ \pi N
\langle X | X \rangle \langle \Psi_{\rm GS} | \Psi_{\rm GS} \rangle }
\]

\beq
\times
\frac{1}{ \omega + i \eta - (E_X - E_{\rm GS} ) } \biggr\} \;\;\; ,
\label{eq22}
\eneq
\noindent
where $|X \rangle$ is an exact one-spinon one-holon eigenstate of
$H_{\rm KY}$,  $|\Phi_{mn} \rangle$ with energy $E_X=E_{mn}$, and $q = 2
\pi k / N$ (below we discuss why only forward propagating states contribute
to Eq.(\ref{eq22})).

Using $A_{\rm h}^{\rm sp,ho} ( \omega , q )$ instead of
$A_{\rm h} ( \omega , q )$ is equivalent to approximating

\[
c_{h_0 \uparrow} | \Psi_{\rm GS} \rangle \approx \Psi_{h_0 h_0} =
\sum_{n=1}^{M+2} \sum_{m=0}^{n-2} h_0^{m-n} p_{mn}(1) \Phi_{mn}  +
\]

\beq
\sum_{n=1}^{M+2} \sum_{m=n-1}^M h_0^{m-n} p_{mn}^{'}(1) \Phi_{mn} \;\;\; .
\label{holedecompositionintoholonandspinon}
\eneq
\noindent
(Eq.(\ref{holedecompositionintoholonandspinon}) basically amounts to neglecting
contributions to $c_{h_0 \uparrow} | \Psi_{\rm GS} \rangle$ coming
from multi-spinon one-holon states.)

Since ${\cal H}_{\rm KY}$ contains the Gutzwiller
projector $P$, its matrix elements between states with at least a
doubly occupied site are zero. Therefore, at half-filling,  $A_h(q, \omega)$
takes contributions only from  forward-propagating hole states. Hence,
using Eqs.(\ref{eq22},\ref{holedecompositionintoholonandspinon}) we obtain:

\[
A_{\rm h}^{\rm sp, ho} ( \omega , q)  =
\Im m \frac{1}{\pi}  \biggl\{ \sum_{l=2}^{M+2 } \sum_{m=0}^{l-2}
\frac{ \delta_{k-m+l} p^2_{ml} ( 1 )}{ \omega + i \eta - (E_{mn} - E_{\rm GS} )
}  +
\]

\beq
 \sum_{l=1}^{M+2 } \sum_{m=l-1}^{M} \frac{ \delta_{k-m+l}
( p^{'}_{ml} )^2 ( 1 ) }{  \omega + i \eta - (E_{mn} - E_{\rm GS} ) }
 \biggr\} \frac{ \langle \Phi_{ml} | \Phi_{ml} \rangle}{
\langle \Psi_{\rm GS} | \Psi_{\rm GS} \rangle } .
\label{eq30}
\eneq
\noindent
Eq.(\ref{eq30}) shows that only the $p_{mn}$'s at $z=1$ determine the
spinon-holon contribution to the hole spectral function. Therefore, the
contribution is completely determined by the spinon-holon interaction.

Let us now analyze the thermodynamic limit of Eq.(\ref{eq30}). In the thermodynamic limit,
 the gamma functions can be approximated by using Stirling's formula
\beq
\Gamma [z] \approx \sqrt{\pi} (z-1)^{z - \frac{1}{2}} e^{-(z-1)} \;\;\; .
\label{stirling}
\eneq
\noindent
From Eqs.(\ref{eq30},\ref{stirling}) we get, in the thermodynamic limit:

\[
A_{\rm h}^{\rm sp,ho } ( q, \omega )  \approx \frac{2}{ \pi  ( M + 1 )} \times
\]

\[
\times \biggl\{ \sum_{l=2}^{M+2 }
\sum_{m=0}^{l-2}
\frac{\delta_{r-m+l}}{{ \omega + i \eta - (E_{mn} - E_{\rm GS} )
}} \sqrt{ \frac{ M - m - \frac{1}{2}}{ m } }
+
\]

\beq
\sum_{l=2}^{M+2 } \sum_{m=l-1}^{M} \frac{\delta_{r-m+l}}{{ \omega + i \eta - (E_{mn} - E_{\rm GS} )
}}
\sqrt{ \frac{ m + \frac{1}{2}}{ M - m }} \biggr\} \;\; .
\label{eq10000}
\eneq
\noindent
(Notice that, in order to stabilize the hole occupation at 1, we had to
introduce a chemical potential $J \pi^2 / 4$, which is added to the
energies $E_{mn}$.)

By defining the auxiliary variables:

\[
q_{\rm sp} = \frac{ 2 \pi}{N} m \;\;\; , \;\;
q_{\rm ho} =  \frac{ 2 \pi}{N} l \;\;\; ,
\]
\noindent
Eq.(\ref{eq10000}), in the thermodynamic limit, may be written in the form
already obtained in \cite{kato}:

\[
A_{\rm h}^{\rm sp\; ho} ( \omega , q )  =
 2 \Im m  \int_0^\pi \frac{ d q_{\rm ho}}{ \pi}  \biggl\{
\int_0^{q_{\rm ho}}  \frac{ d q_{\rm sp}}{ \pi}
\sqrt{ \frac{ \pi-q_{\rm sp}}{ q_{\rm sp}} }
 +
\]

\beq
\int_{q_{\rm ho}}^\pi  \frac{ d q_{\rm sp}}{ \pi}
\sqrt{ \frac{ q_{\rm sp}}{ \pi - q_{\rm sp}}}
 \biggr\} \frac{  \delta ( q - q_{\rm sp} + q_{\rm ho} )}{ \omega -  \mu + i \eta
-E ( q_{\rm sp} , q_{\rm ho} )}  \; .
\label{eq31}
\eneq
\noindent
In the region $0 \le q \le \pi$ the integration of Eq.(\ref{eq31}) gives:

\[
A_{\rm h}^{\rm sp\; ho} ( \omega , q  ) =
\frac{1}{  \pi^2 J q}\sqrt{ \frac{ J [ q + \frac{\pi}{2} ]^2
- \omega }{ \omega - J [ q - \frac{\pi}{2}]^2} }
\]

\beq
\times \Theta \left[ \omega - J [ q - \frac{\pi}{2}]^2 \right]
\Theta \left[ J [ \frac{\pi^2}{4} + q ( \pi - q ) ] - \omega \right] \; .
\label{eq32}
\eneq
\noindent
This formula shows that the spinon-holon probability enhancement
$p_{mn}^2 (1)$ turned into a square-root singularity in
$A_{\rm h}^{\rm sp , ho} ( \omega , q  )$  at the threshold energy for
creation of a spinon-holon pair. Because the spinon-holon joint density of
states is uniform, the main conclusion we trace from our calculation is that
the sharp nonanalytic threshold in $A_{\rm h}^{\rm sp,ho} ( \omega , q  )$
is the direct consequence of spinon-holon interaction.

\section{Conclusions}

In this paper, we have extended the formalism introduced in \cite{us2} to
analyze spinon interaction in the Haldane-Shastry model, to the case where
also charge degrees of freedom are involved. Our formalism allows us
to define a quantum-mechanical real-space representation of the one-spinon
one-holon wavefunction. We construct a Dirac-like equation, whose solution is
the spinon-holon wavefunction in real space coordinates. By means of a
careful study of the real-space one spinon one holon wavefunction, we show
the existence of the spinon-holon interaction and its survival in the
thermodynamic limit. Spinon-holon interaction generates a short-range
enhancement in the probability for a spinon and a holon to be on the same
site. The attraction, however,  is not strong enough to form a spinon-holon
bound state, which would correspond to a Landau's quasihole resonance. This
makes, in the thermodynamic limit, the  hole excitation fully unstable against
decay in one-holon multi-spinon states  and the
quasiparticle peak disappear. Correspondingly, in the thermodynamic
limit, the probability enhancement develops into a square root singularity
followed by a branch cut, which reflects the full instability of the hole
excitation. Hence, by means of a sequence of exact, straightforward steps,
we prove that  spinon-holon attraction is what makes Landau's Fermi liquid
theory break down in 1-d strongly correlated electron systems.

\section*{Acknowledgments}
We would like to thank J. Hennawi, D. Santiago and A. Tagliacozzo for
many interesting discussions and advises about this paper. D. G. acknowledges
the hospitality received from Stanford University during March, 2001, when
this research has been conceived.
This work was supported primarily by the National Science Foundation under
grant No. DMR-9813899. Additional support was provided by the U.S.
 Department of Energy under contract No. DE-AC03-76SF00515.

\appendix

\section{Functions expressed in terms of $\downarrow$-spin coordinates.}

In this Section we prove the formulas that express the 
states of the KY-model in terms of the $\downarrow$-spin coordinates, once
the expression in terms of the $\uparrow$-spin coordinates is known.
The starting point is the following identity ($z_\alpha, z_\beta$ are
$N-th$ roots of the identity):

\beq
\prod_{\alpha: z_\alpha \neq z_\beta} (z_\alpha - z_\beta) =
\lim_{z \rightarrow z_\beta} \frac{ z^N - 1}{ z - z_\beta} =
\frac{N}{ z_\beta} \;\; .
\label{eqa1}
\eneq
\noindent
The ground state wavefunction at filling-1/2 expressed in terms of the
$\uparrow$-spin coordinates is given by:

\beq
\Psi_{GS} ( z_1 , \ldots , z_M ) = \prod_{i<j}^M (z_i - z_j )^2 \prod_t^M z_t \;\; ,
\label{eqa2}
\eneq
\noindent
where $N$ is even and $M = N/2$. Let $\eta_1 , \ldots , \eta_M$ be the
$\downarrow$-spin coordinates. Upon applying Eq.(\ref{eqa1}), we get:

\[
\prod_{i<j}^M (z_i - z_j )^2 \prod_t^M z_t = (-1)^{ \frac{M ( M -1)}{2} }
\prod_{i \neq j}^M (z_i - z_j ) \prod_t^M z_t =
\]

\beq
 (-1)^{ \frac{M ( M + 1)}{2} } \frac{ \prod_t^M z_t \prod_t^M ( \frac{N}{z_t}
)}{ \prod_{ z_j , \eta_i } ( z_j - \eta_i )} =
\prod_{i < j}^M ( \eta_i - \eta_j )^2 \prod_t^M \eta_t \;\; ,
\label{eqa3}
\eneq
\noindent
which proves that $\Psi_{GS} ( z_1 , \ldots , z_M ) = \Psi_{GS} ( \eta_1 ,
\ldots , \eta_M )$.

The one-holon wavefunction is given by:

\beq
\Psi^{\rm ho}_n ( z_1 , \ldots , z_M | h ) = h^n \prod_j^M ( z_j - h )
\prod_{i < j}^M ( z_i - z_j )^2 \prod_t^M z_t \; , 
\label{eqa4}
\eneq
\noindent
where now $N$ is odd, $M = (N-1)/2$ and $h$ is the coordinate of the empty
site.

The same steps as for $\Psi_{GS}$ apply to the one-holon wavefunction. We
have:

\[
h^n \prod_j^M ( z_j - h ) \prod_{i < j}^M ( z_i - z_j )^2 \prod_t^M z_t
\]

\[
= h^n ( -1 )^{ \frac{M ( M + 1)}{2}} \frac{ \prod_t^M z_t \prod_t^M ( \frac{N}{
z_t} )}{ \prod_{ \eta_i , z_j } ( z_j - \eta_i ) }
\]

\beq
= h^n \prod_i^M ( \eta_i - h ) \prod_{i<j}^M ( \eta_i - \eta_j )^2 \prod_t^M
\eta_t \;\; ;
\label{eqa5}
\eneq
\noindent
this proves that $\Psi_h^{\rm ho} ( z_1 , \ldots , z_M |h ) = \Psi_h^{\rm ho}
( \eta_1 , \ldots , \eta_M | h)$.

The one-spinon one-holon state $\Psi_{s}^n ( z_1 , \ldots , z_M | h )$ is given
by:

\[
\Psi_s^n ( z_1 , \ldots , z_M | h ) = 
\]

\beq
h^n \prod_j^M ( z_j - s ) ( z_j - h )
\prod_{i < j}^M ( z_i - z_j )^2 \prod_t^M  z_t \;\; ,
\label{eqa6}
\eneq
\noindent
where $N$ is even, $M=N/2-1$, $s$ is the coordinate of the $\downarrow$-spin
and $h$ is the location of the empty site. As in the previous cases, we have:

\[
h^n \prod_j^M ( z_j - s ) ( z_j - h ) \prod_{i < j}^M
( z_i - z_j )^2 \prod_t^M  z_t
\]

\[
= h^n (-1)^{\frac{M ( M + 1)}{2} } \frac{ \prod_t^M z_t \prod_t^M
( \frac{N}{z_t} )}{\prod_{ \eta_i , z_j } ( z_j - \eta_i )}
\]

\beq
h^n \prod_i^M ( \eta_i - s ) ( \eta_i - h ) \prod_{ i < j}^M ( \eta_i -
\eta_j )^2 \prod_t^M \eta_t  \;\; .
\label{eqa7}
\eneq
\noindent
Eq.(\ref{eqa7}) provides the proof that
 $\Psi_s^n ( z_1 , \ldots , z_M | h ) = \Psi_s^n ( \eta_1 , \ldots ,
\eta_M | h )$.

The last identity we need refers to the case where the spinon and holon are
at the same site, which we had to consider in deriving Eq.(\ref{reverse}).
We have:

\[
\Psi_s^n ( z_1 , \ldots , z_M | s ) = s^n \prod_j^M ( z_j - s)^2
\prod_{i < j}^M ( z_i - z_j )^2 \prod_t^M z_t
\]

\[
= s^n (-1)^{ \frac{ M ( M -1)}{2}} \prod_j^M \frac{ (z_j -s)}{(z_j - h)}
\frac{ ( \prod_t^M z_t)( \prod_t^M ( \frac{N}{ z_t} ) )}{ \prod_{z_j , \eta_i}
( z_j - \eta_i )}
\]

\[
= s^n \prod_j^M \frac{ (z_j -s )}{ ( z_j - h )} \frac{ ( \prod_t^M z_t )
( \prod_t^M ( \frac{N}{ z_t} ))}{ \prod_i^M (\frac{N}{ \eta_i})}
\]

\[
\times
\prod_{ i < j}^M ( \eta_i - \eta_j)^2 \prod_i^M ( \eta_i - s )( \eta_i - h )
\]

\beq
= - ( \frac{s}{h} )^{n-1} h^n \prod_i^M ( \eta_i - h)^2 \prod_{ i < j}
( \eta_i - \eta_j )^2  \prod_t^M \eta_t  \; .
\label{eqa8}
\eneq
\noindent
Eq.(\ref{eqa8}) proves the identity

\[
\Psi_s^n ( z_1 , \ldots , z_M | s ) = - \left( \frac{s}{h} \right)^{n-1}
\Psi_h^n ( \eta_1 , \ldots , \eta_M | h ) \;\;\; .
\]

\section{The norm of one-holon wavefunction.}

In this Appendix we discuss in detail the calculation of the norm of the
negative-energy one-holon states,
$\langle \Psi_n^{\rm ho} | \Psi_n^{\rm ho} \rangle$. 

\beq
\langle \Psi_n^{\rm ho} | \Psi_n^{\rm ho} \rangle =
\sum_{ z_1 , \ldots , z_M , h} | \Psi_n^{\rm ho} ( z_1 ,
\ldots , z_M | h ) |^2 \;\; .
\label{eqbb1}
\eneq
\noindent
Let $P_m ( z_1 , \ldots , z_M )$ be the $m-th$ degree symmetric polynomial
in $z_1 , \ldots, z_M$. One obtains \cite{us2}:

\[
\Psi_n^{\rm ho} ( z_1 , \ldots , z_M | h )
\]

\beq
= \sum_{m=0}^M h^{m+n}
P_m ( z_1 , \ldots , z_M ) \prod_{i < j}^M (z_i - z_j )^2 \;\;\; .
\label{eqbb2}
\eneq
\noindent
From Eqs.(\ref{eqbb1},\ref{eqbb2}), we derive:

\[
\langle \Psi_n^{\rm ho} | \Psi_n^{\rm ho} \rangle  =
N \sum_{m=0}^M \sum_{ z_1 , \ldots , z_M} | P_m ( z_1 , \ldots ,
z_M ) |^2 \prod_{i < j}^M | z_i - z_j |^4
\]

\beq
= N \sum_{m=0}^M \langle \Psi_m^{\rm sp} | \Psi_m^{\rm sp} \rangle \;\;\; .
\label{eqb3}
\eneq
\noindent
The norm of one-spinon wavefunctions, $\langle \Psi_m^{\rm sp} |
\Psi_m^{\rm sp} \rangle $, has been derived in Ref.\cite{us2}, where it has
been shown that:

\[
\langle \Psi_m^{\rm sp} | \Psi_m^{\rm sp} \rangle
\]

\beq
= \frac{ \Gamma [ M + 1] }{
\Gamma [ M + \frac{1}{2} ] } \frac{ \Gamma [ m + \frac{1}{2} ] \Gamma [ M - m
+ \frac{1}{2} ] }{ \Gamma [ m + 1 ] \Gamma [ M - m + 1 ]}
N^M \frac{ ( 2 M)!}{ 2^M } .
\label{eqb4}
\eneq
\noindent
Therefore, we can re-write Eq.(\ref{eqb3}) as:

\beq
\langle \Psi_n^{\rm ho} | \Psi_n^{\rm ho} \rangle =
\Gamma [ \frac{1}{2} ]  \frac{ ( 2 M)!}{ 2^M } g_{M 0} ( 1 ) \;\; ,
\label{eqb5}
\eneq
\noindent
where the polynomial $g_{mn} (z)$ is the two-spinon relative
wavefunction, denoted by $p_{mn} ( z )$ in Ref.\cite{us2}:

\[
g_{mn} ( z ) = \frac{ \Gamma [ m - n + 1 ]}{ \Gamma [ \frac{1}{2} ]
\Gamma [ m - n + \frac{1}{2} ] } 
\]

\beq
\times
\sum_{k=0}^{m-n} \frac{ \Gamma [ k +
\frac{1}{2} ] \Gamma [ m - n - k + \frac{1}{2} ] }{ \Gamma [ k + 1 ]
\Gamma [ m - n - k + 1] } z^k \;\;\; .
\label{eqb6}
\eneq
\noindent
By means of generic properties of hypergeometric functions \cite{abram}, one
gets:

\beq
g_{mn} ( 1 ) = \frac{ \Gamma [ \frac{1}{2} ] \Gamma [ m - n + 1 ]}{
\Gamma [ m - n + \frac{1}{2} ] } \;\;\; .
\label{eqb7}
\eneq
\noindent
Therefore, we obtain:

\beq
\langle \Psi_n^{\rm ho} | \Psi_n^{\rm ho} \rangle = N^{M+1}
\frac{ ( 2 M)!}{ 2^M } \frac{ \Gamma [ \frac{1}{2} ] \Gamma [ M +  1 ]}{
\Gamma [ M + \frac{1}{2} ] } \;\; ; \forall n \; .
\label{eqb8}
\eneq
\noindent

\section{The norm of one-spinon one-holon energy eigenstates.}

In this section we generalize the recursion procedure introduced in
Ref.\cite{us1,us2} to calculate the norm of one-spinon and two-spinon
wavefunction to the calculation of the norm of one-spinon one-holon
energy eigenstates. As in the calculation in Refs.\cite{us1,us2},
the key operator is given by $e_1 ( z_1 , \ldots , z_M )$, defined as:

\beq
e_1 ( z_1 , \ldots , z_M ) = z_1 + \ldots + z_M \;\;\; .
\label{eqc1}
\eneq
\noindent
The state for one holon and one spinon localized at $s$ is given by:

\beq
\Psi_s^n (z_1 , \ldots , z_M ) = \Phi_s^n ( z_1 , \ldots , z_M | h )
\Psi_{GS} ( z_1 , \ldots , z_M ) \; ,
\label{eqc2}
\eneq
\noindent
where:

\[
 \Phi_s^n ( z_1 , \ldots , z_M | h ) = h^n \prod_j^M (z_j - s)( z_j - h) \;\; .
\]
\noindent
On $\Psi_s^n (z_1 , \ldots , z_M )$, $e_1$ acts as a ladder operator, as we
are going to show next.

In order to work out the action of $H_{\rm KY}$ on holon eigenstate, we
splitted it into five terms: $H_{\rm KY} / \frac{J}{2} ( \frac{ 2 \pi}{N} )^2
= h_S^T + h_S^V + h_N + h_Q^\downarrow + h_Q^\uparrow$. Among those five
terms, the only ones that do not commute with $e_1$ are the spin
exchange operator $h_S^T$ and the $\uparrow$-spin charge  propagation
operator $h_Q^\uparrow$. On the state $\Psi_s^n$, $h_S^T$ is realized as:

\[
h_S^T \Psi_s^n = \Psi_{GS} \biggl\{ const. + \frac{1}{2} \sum_i^M z_i^2
\frac{ \partial^2}{\partial z_i^2}
\]

\beq
 + 2 \sum_{i \neq j}^M \frac{ z_i^2}{
z_i - z_j} \frac{ \partial}{ \partial z_i} - \frac{N-3}{2}  \sum_i^M z_i^2
\frac{ \partial}{ \partial z_i} \biggr\} \Phi_s^n \; .
\label{eqc3}
\eneq
\noindent
From Eq.(\ref{eqc3}), we derive that:

\[
[ h_S^T e_1 \Phi_s^n \Psi_{GS} ] - e_1 \Phi_s^n [ h_S^T \Psi_{GS} ]
\]

\[
=
\Psi_{GS} \biggl\{ \left[ M - \frac{3}{2} \right] e_1 \Phi_s^n +
\sum_i^M z_i^2 \frac{ \partial}{ \partial z_i} \Phi_s^n \biggr\}
\]

\[
= \Psi_{GS} \biggl\{ ( M + \frac{1}{2} ) e_1 \Phi_s^n + M ( s + h ) \Phi_s^n
\]

\beq
- s^2 \frac{ \partial}{ \partial s} \Phi_s^n
- h^{n+2} \frac{ \partial}{\partial h} \left( \frac{ \Phi_s^n}{ h^n} \right)
\biggr\} \;\;\; .
\label{eqc4}
\eneq
\noindent
In the sense clarified by Eq.(\ref{eqc4}), $e_1$ commutes with $h_S^V$,
$h_N$ and $h_Q^\downarrow$. On the other hand, it does not commute with
$h_Q^\uparrow$. Indeed, since, in order to derive the action of $h_Q^\uparrow$
on $\Psi_s^n$ we have to express the state in terms of the $\downarrow$-spin
coordinates, we must do the same with $e_1$. In order to do so, we notice that
$\{z_j \}$, $\{ \eta_j \}$, $h$ and $s$ taken all together are the set of the
$N$ $n-th$ roots of 1. Therefore, we obtain:

\[
z_1 + \ldots + z_M + \eta_1 + \ldots + \eta_M + h + s = 0 \;\; ,
\]
\noindent
which implies:

\beq
e_1 ( z_1 , \ldots , z_M ) = - e_1 ( \eta_1 , \ldots , \eta_M ) - h - s \; .
\label{eqc5}
\eneq
\noindent
The action of $h_Q^\uparrow$ on $e_1 \Psi_s^n$ gives:

\[
[ h_Q^\uparrow e_1 ] \Psi_s^n = \Psi_{GS} \biggl\{ \left[
\frac{N^2-1}{12} + \frac{n (n-N)}{2} \right] e_1 \Phi_s^n ( \{ \eta \} | h )
\]

\[
- \left[ \frac{N-1}{2} - n \right] h^{n+1} \frac{ \partial}{ \partial h}
\left[ ( - h - s - \sum_j \eta_j ) \frac{ \Phi_s^n ( \{ \eta \} | h)  }{ h^n}
\right]
\]

\[
+ \frac{1}{2} h^{n+2} \frac{ \partial^2}{ \partial h^2}
\left[ ( - h - s - \sum_j^M \eta_j ) \frac{ \Phi_s^n (
\{ \eta \} | h ) }{ h^n} \right]
\]

\[
+ \frac{sh}{ (s-h)^2} \left[ - \sum_j \eta_j - 2 s \right] \Phi_s^n (
\{ \eta \}  | s ) \biggr\}
\]

\[
= \Psi_{GS} e_1 [ h_Q^\uparrow \Phi_s^n ]
( \eta_1 , \ldots , \eta_M | h
)
\]

\[
+ \left[ \frac{N-1}{2} - n \right] h \Phi_n^s
( \eta_1 , \ldots , \eta_M | h)
\]

\[
-  h^{n+2} \frac{ \partial}{\partial h}
\left( \frac{\Phi_s^n  ( \eta_1 , \ldots , \eta_M | h)}{h^n} \right)
\]

\beq
+ \frac{h}{h-s} \Phi_s^{n+1}  ( \eta_1 , \ldots , \eta_M | s) \;\;\; .
\label{eqc7}
\eneq
\noindent
Eq.(\ref{eqc7}) yields the result:

\[
[ h_Q^\uparrow , e_1 ] \Psi_s^n =
\]

\[
\Psi_{GS} \biggl\{ h \frac{h}{h-s} \biggl[ \Phi_s^n
( z_1 , \ldots , z_M | h )
- \left( \frac{s}{h} \right)^n  \Phi_h^n ( z_1 , \ldots , z_M | h ) \biggr]
\]

\beq
+ h^{n+2} \frac{ \partial}{ \partial h} \left( \frac{ \Phi_s^n ( z_1 ,
\ldots , z_M | h )}{ h^n} \right) - n h \Phi_s^n
( z_1 , \ldots , z_M | h )  \biggr\} .
\label{eqc8}
\eneq
\noindent
Upon summing Eq.(\ref{eqc4}) and Eq.(\ref{eqc8}), we derive the basic relation
we need in order to work out the recursion relations:

\[
[ \frac{H_{\rm KY}}{ \frac{J}{2} \left( \frac{2\pi}{N} \right)^2} , e_1 ]
\Psi_s^n =   ( M + \frac{1}{2} ) e_1 \Psi_s^n + M (s+h)
 \Psi_s^n
\]

\[
 - s^2 \frac{ \partial}{ \partial s} \Psi_s^n + h \frac{h}{h-s}
\sum_{m=0}^M \left[ s^m - \left( \frac{s}{h} \right)^n h^m \right] \Psi_{mn}
\]

\beq
- n h \Psi_s^n \;\;\; ,
\label{eqc9}
\eneq
\noindent
where we have introduced the one-spinon one-holon plane waves $\Psi_{mn}$
defined in Section VI-B.

In order to further manipulate term in Eq.(\ref{eqc9}), let us consider, now,
the identity:

\[
h \frac{h}{h-s}\sum_s \frac{s^{-k}}{N}
\sum_{m=0}^M \left[ s^m - \left( \frac{s}{h} \right)^n h^m \right] \Psi_{mn}
\]

\[
= \sum_{m=0}^M \Psi_{mn} \frac{ h^{1+m-k}}{N} \sum_{ \{ s/h \} } \left[ \frac{
(\frac{s}{h})^{n-k}}{\frac{s}{h}-1} - \frac{(
\frac{s}{h})^{m-k}}{\frac{s}{h}-1} \right]
\]

\[
= \frac{1}{N} \sum_{m=0}^{n-1} \Psi_{mn} h^{1+m-k} \sum_{ \{ s/h \} }
\sum_{r=0}^{n-m-1}  \left(\frac{s}{h} \right)^{m+r-k}
\]

\beq
- \frac{1}{N} \sum_{m=n+1}^M \Psi_{mn}  h^{1+m-k} \sum_{ \{ s/h \} }
\sum_{r=0}^{m-n-1} \left( \frac{s}{h} \right)^{n+r-k}  .
\label{eqc10}
\eneq
\noindent
(The symbol $\sum_{ \{ s /h \} }$ means that we have to sum over
$s/h \in S^N$).

Suppose, now, $n-k+1 <0$. In this case, the sum in Eq.(\ref{eqc10}) over
$m$ from $n+1$ to $M$ will  give 0, while the second sum will be reduced to:

\[
\sum_{m=0}^k \Psi_{mn} h^{1+m-k} \sum_{ \{ s/h \} } \sum_{r=0}^{n-m-1}
\left(\frac{s}{h} \right)^{m+r-k} \;\;\; .
\]
\noindent
On the other hand, as $k-n+1 \geq 0$, the only nonzero term in the sum will
be given by:

\[
-\sum_{m=k+1}^M \Psi_{mn} h^{1+m-k} \sum_{ \{ s/h \} } \sum_{r=0}^{m-n-1}
\left(\frac{s}{h} \right)^{n-k+r} \;\;\; .
\]
\noindent
In order to write down the action of $ [ H_{\rm KY} , e_1 ]$ on the plane-wave
states $\Psi_{k n}$, we multiply both members of Eq.(\ref{eqc8}) by $s^{-k}$
and sum over $s$. By using Eq.(\ref{eqc10}), it is straightforward to derive
the following equations:

\begin{itemize}

\item If $k-n+1 < 0$:

\[
[ \frac{H_{\rm KY}}{ \frac{J}{2} ( \frac{2\pi}{N} )^2 }  , e_1 ] \Psi_{k n} =
( M + \frac{1}{2} ) e_1 \Psi_{k n}
\]

\[
 + ( M-k+1) \Psi_{k-1,n} + (M-n+1) h \Psi_{kn}
\]

\beq
 + \sum_{m=0}^{k-1}
\Psi_{mn} \frac{h^{1+m-k}}{N}
\sum_{ \{ s/h \} }  \sum_{r=0}^{n-m-1} \left( \frac{s}{h} \right) ^{m-k+r} .
\eneq

\item If $k-n+1 \geq 0$:

\[
[ \frac{H_{\rm KY}}{ \frac{J}{2} ( \frac{2\pi}{N} )^2 }  , e_1 ]
\Psi_{k n} =  ( M + \frac{1}{2} ) e_1 \Psi_{kn}
\]

\[
+ ( M-k+1) \Psi_{k-1,n} + (M-n)  h \Psi_{kn}
\]

\beq
- \sum_{m=k+1}^M \Psi_{mn}
\frac{h^{1+m-k}}{N} \sum_{ \{ s/h \} } \sum_{r=0}^{m-n-1}
\left( \frac{s}{h} \right)^{n-k+r} .
\label{eqc11}
\eneq
\end{itemize}

Let us now work out the basic recursion relation in both cases.

\begin{itemize}

\item Case $k-n+1 < 0$.

In this case energy eigenstates are given by:

\[
\Phi_{kn} = \sum_{\ell=0}^k a_\ell \Phi_{k-\ell,n-\ell} \;\;\; .
\]
\noindent
(see Section VI-B for the definition of the coefficients $a_j$).

Therefore, we have:

\[
[ \frac{H_{\rm KY}}{ \frac{J}{2} ( \frac{ 2 \pi}{N} )^2 } , e_1 ]
\Phi_{kn} = \sum_{\ell=0}^k a_\ell [  \frac{H_{\rm KY}}{ \frac{J}{2}
( \frac{ 2 \pi}{N} )^2 } , e_1 ] \Psi_{k-\ell,n-\ell}
\]

\[
=\sum_{\ell=0}^k a_\ell ( M + \frac{1}{2} ) e_1  \Psi_{k-\ell,n-\ell}
\]

\[
+ \sum_{\ell=0}^{k-1} a_\ell (M-k+\ell+1 ) \Psi_{k-1-\ell,n-\ell}
\]

\[
+ \sum_{\ell=0}^k a_\ell (M-n+\ell+1)  \Psi_{k-\ell,n-\ell+1}
\]

\[
+ \sum_{\ell=0}^{k-1} a_\ell \sum_{m=0}^{k-\ell-1} \Psi_{m,n+1+m-k}
\]

\beq
\times
 \left[ \frac{1}{N}  \sum_{ \{ s/h \} } \sum_{r=0}^{n-m-1} \left( \frac{s}{h}
\right)^{m-k+r} \right] \;\; .
\label{eqcc11}
\eneq

\noindent
From Eq.(\ref{eqcc11}), we get:

\[
( E_{k-1,n} - E_{kn} - M - \frac{1}{2} ) \langle \Phi_{k-1,n} | e_1 | \Phi_{
kn} \rangle
\]

\[
= \sum_{\ell=0}^{k-1} a_\ell (M-k+\ell+1) \langle \Phi_{k-1,n} |
\Psi_{k-1-\ell , n-\ell } \rangle +
\]

\[
\sum_{\ell=0}^{k} a_\ell (M-n+\ell+1) \langle \Phi_{k-1,n} |
\Psi_{k-\ell,n+1-\ell} \rangle
+  \langle \Phi_{k-1,n} |\Psi_{k-1 , n } \rangle
\]

\[
= [ M-k+2 + b_1 (M-n+1) + a_1 (M-n+2) ] \langle \Phi_{k-1,n} |\Phi_{k-1,n}
\rangle
\]

\beq
 = ( M - k + \frac{3}{2} )  \langle \Phi_{k-1,n} |\Phi_{k-1,n} \rangle \;\; .
\label{eqc12}
\eneq
\noindent
(notice that, from their definition, we have $a_1 = - 1/2$, $b_1 = 1/2$).
Eq.(\ref{eqc12}) may be recast in the following compact form:

\beq
\frac{ \langle \Phi_{k-1, n } | e_1 | \Phi_{kn} \rangle}{\langle
\Phi_{k-1,n} | \Phi_{k-1,n} \rangle }  =
- \frac{ M - k +\frac{3}{2}}{ 2 ( M-k+1)}
\label{eqc13}
\eneq

\item Case $k-n+1 \geq 0$.

In this case we have:

\[
\Phi_{kn} = \sum_{\ell=0}^{M-k} a_\ell \Phi_{k+\ell, n+\ell} \;\;\; .
\]
\noindent

Eq.(\ref{eqcc11}) now takes the form:

\[
[ \frac{H_{\rm KY}}{\frac{J}{2} ( \frac{ 2 \pi}{N} )^2} , e_1 ] \Phi_{kn}
\]

\[
=\sum_{\ell=0}^{M-k} a_\ell ( M + \frac{1}{2} ) e_1 \Psi_{k+\ell,n+\ell}
\]

\[
+ \sum_{\ell=0}^{M-k} a_\ell (M-k-\ell+1) \Psi_{k+\ell-1,n+\ell}
\]

\[
+ \sum_{\ell=0}^{M-k} a_\ell (M-n-\ell) \Psi_{k+\ell,n+1+\ell}
\]

\[
-\sum_{\ell=0}^{M-k} a_\ell \sum_{m=k+\ell+1}^M \Psi_{m,n+1+m-k}
\]

\beq
\times \frac{1}{N} \sum_{\{ s/h \} } \left[
\sum_{r=0}^{m-n-1} \left( \frac{s}{h} \right)^{n-k+r} \right] \;\; . 
\label{eqcc13}
\eneq
\noindent
From Eq.(\ref{eqcc13}) and by working exactly as in the previous case, we get
the identity:

\[
(E_{k-1,n} - E_{kn} - M -\frac{1}{2} ) \langle \Phi_{k-1,n} | e_1
| \Phi_{kn} \rangle
\]

\beq
= (M-k+1 ) \langle \Phi_{k-1,n} | \Phi_{k-1,n} \rangle \;\; ,
\label{eqc14}
\eneq
\noindent
which yields the relation:

\beq
\frac{ \langle \Phi_{k-1, n} | e_1 | \Phi_{kn} \rangle}{ \langle \Phi_{k-1,n}
| \Phi_{ k-1,n} \rangle} = - \frac{ M - k + 1}{ 2 ( M - k + \frac{1}{2} )} .
\label{eqc15}
\eneq
\noindent
In order to complete the recursion procedure, we need one more induction
relation that is derived by inserting in the product $\langle \Phi_{ab} |
e_1 | \Phi_{cd} \rangle$ the product $e_M e_M^* = 1$, where:

\[
e_M ( z_1 , \ldots , z_M ) = z_1 \cdot \ldots \cdot z_M \;\;\; .
\]
\noindent
From the definition of scalar product, Eq.(\ref{nor1}), it is straightforward
to show  that:

\end{itemize}

\[
\langle \Phi_{ab} | e_1 | \Phi_{cd} \rangle =
\langle [ ( e_M )^2 \Phi_{ab} ]  | e_1 | [  ( e_M )^2 \Phi_{cd}]  \rangle
\]

\beq
= \langle \Phi_{M-c, M-d} | e_1 | \Phi_{M-a , M-b} \rangle \;\;\; .
\label{eqc16}
\eneq
\noindent
By applying Eq.(\ref{eqc16}) to Eq.(\ref{eqc12})  and by using
Eq.(\ref{eqc15}), we obtain:

\[
 \langle \Phi_{k-1,n} | e_1 | \Phi_{kn} \rangle = \langle \Phi_{M-k,M-n} |
e_1 | \Phi_{M-k+1,M-n} \rangle
\]

\beq
= - \frac{ k}{ 2 ( k - \frac{1}{2}) } \langle \Phi_{kn} | \Phi_{kn} \rangle \;\;\; .
\label{eqc17}
\eneq
\noindent
By putting together Eq.(\ref{eqcc13}) and Eq.(\ref{eqc17}) one finally
obtains:

\beq
\frac{ \langle \Phi_{kn} | \Phi_{kn} \rangle}{ \langle \Phi_{k-1,n} |
\Phi_{k-1,n} \rangle} = \frac{ ( k - \frac{1}{2} ) ( M - k + \frac{3}{2} )}{
k (M-k+1)} \;\; ,
\eneq
\noindent
that provides the formula for the norm of the one-spinon one-holon energy
eigenstates in the case $k-n+1 < 0$:

\beq
\langle \Phi_{kn} | \Phi_{kn} \rangle = N^{M+1} \frac{ (2M)!}{2^M}
( M + \frac{1}{2} ) \frac{ \Gamma [ M - k + 1] \Gamma [ k +
\frac{1}{2}]}{ \Gamma [ M - k + \frac{3}{2} ] \Gamma [ k + 1 ]} \; .
\eneq
\noindent
In the complementary case, $k-n+1 \geq 0$, we may follow the same step to
prove that:

\[
\langle \Phi_{kn} | \Phi_{kn} \rangle = \langle \Phi_{M-k,M-n} |
\Phi_{M-k,M-n} \rangle
\]

\beq
=  N^{M+1} \frac{ (2M)!}{2^M} ( M + \frac{1}{2} )
\frac{ \Gamma [ M-k+\frac{1}{2} ] \Gamma [ k + 1 ] }{ \Gamma [ M - k +1 ]
\Gamma [ k + \frac{3}{2} ] } .
\eneq

\section{Solution of the equation of motion for the one-spinon one-holon
wavefunction}

In this Appendix we will derive the solution to the equation of motion
for the relative coordinate part of the spinon-holon wavefunctions,
$p_{mn} (z)$, $p^{'}_{mn} (z)$. In order to do so, let us consider
first the case $m - n + 1 < 0$, in which we express the solution to
Eq.(\ref{eq18}) as a power series of $z$:

\beq
p_{mn} ( z ) = \sum_k a_k z^k \;\;\; .
\label{eqd2}
\eneq
\noindent
From Eq.(\ref{eqd2}), we get the following equation for the
coefficients $a_k$:

\beq
- 2 \sum_k  k  a_{k+1} z^k + \sum_k ( 2 k -1 ) a_k z^k -
z^{M-m} \sum_k a_k = 0 .
\label{eqd3}
\eneq
\noindent
As $k \leq M-m$, the following recursion relation between the $a_k$'s
holds:

\beq
\frac{ a_{k+1}}{ a_k} = \frac{ k + \frac{1}{2} }{ k + 1} \;\;\; .
\label{eqd4}
\eneq
\noindent
Eq.(\ref{eqd4}) is satisfied by

\beq
a_k = \frac{ \Gamma [ k + \frac{1}{2} ] }{ \Gamma [ \frac{1}{2} ]
\Gamma [ k + 1 ]} \;\;\; .
\label{eqd5}
\eneq
\noindent
To calculate $a_{M-m+1}$, we need the following identity, valid
for any positive integer $R$ ($C_0$ is a closed path centered at $z=0$):

\[
\sum_{k = 0}^R \frac{ \Gamma [ k + \frac{1}{2} ] }{ \Gamma [ k + 1 ] } =
\sum_{k=0}^R \oint_{C_0} \frac{ d z }{ 2 \pi i } \frac{1}{z^{k+1}}
\frac{1}{ \sqrt{ 1 - z}} =
\]

\[
\oint_{C_0} \frac{ d z}{ 2 \pi i z^{R+1} } \frac{ z^{R+1} - 1}{z - 1}
\frac{1}{ \sqrt{ 1 - z}} =
\]

\beq
\oint_{C_0} \frac{ d z}{ z^{R+1} 2 \pi i } \frac{1}{ ( 1 - z)^\frac{3}{2} }
=2 \frac{ \Gamma [ R + \frac{1}{2} ] }{ \Gamma [ R + 1 ]} ( R + \frac{1}{2} ) \; .
\label{eqd6}
\eneq
\noindent
Since the recursion relation for $a_{M-m}$ is:

\[
- 2 ( M - m  ) a_{ M -m} + 2 ( M- m - \frac{1}{2} ) a_{M-m-1}
\]

\beq
- \sum_k a_k = 0  \;\;\; ,
\label{eqd7}
\eneq

\noindent
Eq.(\ref{eqd6}) implies $a_{M-m} = 0$ and:

\beq
p_{mn} ( z ) = \sum_{ k = 0}^{M-m-1} \frac{ \Gamma [ k + \frac{1}{2} ] }{
\Gamma [ \frac{1}{2} ] \Gamma [ k + 1 ] } z^k  \;\;\; .
\label{eqd8}
\eneq
\noindent
Eq.(\ref{eqd8}) provides the formula for $p_{mn} (z)$ used
throughout the paper.

The same steps followed in the case  $m -n +1 < 0$ allow us to find
the spinon-holon wavefunction in the case $m -n +1 \geq0$, 
provided the coordinate $z$ is substituted with  $\xi = 1/z$.

\end{document}